\documentclass[twocolumn]{aastex631}

\usepackage{graphicx}
\usepackage{longtable}
\usepackage{ulem}
\usepackage{amsmath}	% Advanced maths commands
\usepackage{amssymb}	% Extra maths symbols

\def\oiii{\rm{[O \sc{iii}]}}

\begin{document}

\title{Powerful outflows of compact radio galaxies}

%%%%% AUTHORS - PLACE YOUR OWN PACKAGES HERE %%%%%

% Bárbara
\newcommand{\orcidauthorA}{0000-0002-7205-6332}

%Alberto
\newcommand{\orcidauthorB}{0000-0002-7608-6109}

%Marcos 
%https://orcid.org/0000-0002-7989-9041
\newcommand{\orcidauthorC}{0000-0002-7865-3971}

%Stavros
\newcommand{\orcidauthorD}{0000-0003-1351-7204}

%Panda
\newcommand{\orcidauthorX}{0000-0002-5854-7426}

\newcommand{\ha}{H\,$\alpha$} 
\newcommand{\hb}{H\,$\beta$}
\newcommand{\hc}{H\,$\gamma$}

\newcommand{\argoniii}{[Ar\,{\sc iii}]}
\newcommand{\helium}{He\,{\sc i}}
\newcommand{\heliumb}{He\,{\sc ii}}
\newcommand{\ariii}{[Ar\,{\sc iii}]~7136~\AA}
\newcommand{\nia}{[N\,{\sc i}]~5199~\AA}
\newcommand{\oi}{[O\,{\sc i}]~6300~\AA}
\newcommand{\oiia}{[O\,{\sc ii}]~7320~\AA}
\newcommand{\oiib}{[O\,{\sc ii}]~7330~\AA}
\newcommand{\niia}{[N\,{\sc ii}]~5755~\AA}
\newcommand{\niib}{[N\,{\sc ii}]~6584~\AA}
\newcommand{\niic}{[N\,{\sc ii}]~6548~\AA} 
\newcommand{\siia}{[S\,{\sc ii}]~6717~\AA}
\newcommand{\siib}{[S\,{\sc ii}]~6731~\AA}
\newcommand{\cliiia}{[Cl\,{\sc iii}]~5517~\AA}
\newcommand{\cliiib}{[Cl\,{\sc iii}]~5538~\AA} 
\newcommand{\NII}{[N\,{\sc ii}]~6548~\&~6584~\AA}
\newcommand{\oiiia}{[O\,{\sc iii}]~4959~\AA} 
\newcommand{\oiiib}{[O\,{\sc iii}]~5007~\AA} 
\newcommand{\sii}{[S\,{\sc ii}]~6717 \& 6731~\AA}
\newcommand{\ariv}{[Ar\,{\sc iv}]~4711 \& 4740~\AA}
\newcommand{\siiia}{[S\,{\sc iii}]~6312~\AA} 
\newcommand{\siiib}{[S\,{\sc iii}]~9069~\AA} 
\newcommand{\nitrogen}{[N\,{\sc ii}]}
\newcommand{\nitrogena}{[N\,{\sc i}]}
\newcommand{\oxygeniii}{[O\,{\sc iii}]}
\newcommand{\oxygeniiib}{O\,{\sc iii}}
\newcommand{\oxygeni}{[O\,{\sc i}]}
\newcommand{\oxygenii}{[O\,{\sc ii}]}
\newcommand{\sulfur}{[S\,{\sc iii}]}
\newcommand{\sulfurt}{[S\,{\sc ii}]}
\newcommand{\chloro}{[Cl\,{\sc iii}]}
\newcommand{\chloroii}{[Cl\,{\sc ii}]}

\def\aap{A\&A\/}
\def\apjs{ApJS}
\def\apj{ApJ}
\def\pasp{PASP}
\def\aj{AJ}
\def\nat{Nat}
\def\apjl{ApJL}
\def\apss{ApSS}
\def\mnras{MNRAS}
\def\pasj{PASJ}
\def\nar{NAR}
\def\araa{ARA\&Ap}
\def\rmxaa{RevMexA\&Ap}

\correspondingauthor{Bárbara L. Miranda Marques}
\email{barbaralmmarques@gmail.com}

\author[0000-0002-7205-6332]{Bárbara L. Miranda Marques}
\affiliation{Instituto Nacional de Pesquisas Espaciais, Av. dos Astronautas 1758, São José dos Campos, 12227-010, SP, Brazil}

\author[0000-0002-7608-6109]{Alberto Rodríguez-Ardila}
\affiliation{Laboratório Nacional de Astrofísica, Rua dos Estados Unidos, 154, Itajubá/MG, 37504-364, Brazil}
\affiliation{Instituto Nacional de Pesquisas Espaciais, Av. dos Astronautas 1758, São José dos Campos, 12227-010, SP, Brazil}
\affiliation{Observatório Nacional, Rua General José Cristino 77, CEP 20921-400, São
Cristóvão, Rio de Janeiro, RJ, Brazil}

\author[0000-0002-7865-3971]{Marcos A. Fonseca-Faria}
\affiliation{Laboratório Nacional de Astrofísica, Rua dos Estados Unidos, 154, Itajubá/MG, 37504-364, Brazil}

\author[0000-0002-5854-7426]{Swayamtrupta Panda}
%\altaffiliation{CNPq Fellow}
\altaffiliation{Gemini Science Fellow}
\affiliation{Laboratório Nacional de Astrofísica, Rua dos Estados Unidos, 154, Itajubá/MG, 37504-364, Brazil}
\affiliation{International Gemini Observatory/NSF NOIRLab, Casilla 603, La Serena, Chile}

\begin{abstract}

Gigahertz Peaked Spectrum (GPS) and Compact Steep Spectrum (CSS) sources are compact radio galaxies (RGs), with jets extending up to 20 kpc and ages $<$~10$^3$ years. They are considered to evolve to Fanaroff-Riley RGs, but the real scenario to explain the compact sources remains unsolved. The young compact jets make GPS/CSS ideal for studying feedback in the nuclear region of AGNs because the jets are just starting to leave this region. Numerical simulations and jet power estimates suggest that compact sources can drive outflows on scales several times larger than the radio source itself, but the lack of suitable data limits comparisons between theory and observation. We carried out an optical spectroscopic study of 82 CSS/GPS with SDSS-DR12 data to investigate the influence of compact jets in the gas. We found outflowing gas components in the \rm{[O \sc{iii}]}~$\lambda5007$ emission lines in half of our sample. The kinetic energy of the outflowing gas in compact sources is comparable to that observed in extended RGs, indicating that the compact jets can drive powerful outflows similar to those in FR RGs. The observed anti-correlation between the kinetic power of the outflow and the radio luminosity suggests an interaction between the young jet and the interstellar medium (ISM). This finding provides significant observational support for previous simulations of jet-ISM interactions and supports the evolutionary scenario for RGs. However, the lack of sources with high kinetic efficiency indicates that some compact galaxies may be frustrated sources.

\end{abstract}

\keywords{galaxies: active -- galaxies: individual: CSS/GPS -- galaxies: jets -- ISM: jets and outflow -- radio lines: galaxies}

\section{Introduction} \label{sec:intro}

Gigahertz Peaked Spectrum (GPS) and Compact Steep Spectrum (CSS) sources are compact radio galaxies (RGs), with jets ranging from a few parsecs to $\sim$20~kpc in projected linear size and ages $t_{\rm jet} <10^{3}$ years \citep{ODea21}. They contrast dramatically with well-known RGs, where the jets can reach linear extensions of hundreds of kiloparsecs \citep{hardcastle98} or even a few Mpc \citep{Machalski08}. It is estimated that  10\% of the RGs are CSS and 30\% are GPS \citep{ODea98b}. Radio luminosities are above 10$^{25}$~W\,Hz$^{-1}$ at 1.6~GHz and less luminous CSS/GPS have emission with flux density less than 1~Jy.
GPS/CSS derived their name from the characteristic peak in the radio spectrum. In GPS, it is located close to 1~GHz, while in CSS the peak occurs at frequencies smaller than 400~MHz. It is believed that the peak is due either to synchrotron self-absorption or to free-free absorption. Sources with radio spectral peak below $\sim$400~MHz are classified as Megahertz Peaked Spectrum \citep[MPS,][]{Coppejans16}, whose megahertz peak is believed to be a result of a shifted higher frequency peak from CSS and/or GPS and HFP (High-Frequency Peakers, with $\nu > 5$~GHz).

There are three main hypotheses to explain the compact nature of CSS/GPS jets: they are young objects; the jets are confined by a dense interstellar medium; or they are intermittent sources. Each of these hypotheses has observational support for individual objects \citep{Callingham15,ODea91}, but the real nature of the small jets of CSS and GPS is still unknown. 
In a general scenario of the evolution of RGs, it is believed that GPS are the initial stage, evolving to a CSS on their way to becoming a giant RG \citep[Fanaroff-Riley I or II, FR~I or FR~II, respectively,][]{Fanti95,An12}. However, there is an excess of compact RGs relative to the extended ones, suggesting that not all CSS/GPS reach the FR~I/FR~II stage \citep{Turner15,Kunert-Bajraszewska16}. Studies suggest that the strong interaction of young jets with the dense environments precludes their propagation further out from the circumnuclear region, becoming frustrated objects \citep{ODea91}. In contrast to the idea of evolution, other studies suggest that CSS/GPS would have transient nuclear activity, with intermittent jets \citep{Callingham15}.

The presence of young compact jets makes GPS/CSS galaxies perfect targets to study feedback in the nuclear region of AGNs. This is because the jets are just starting to leave the central region so they are expected to interact with the surrounding gas when channeling through the ISM \citep{Duggal21}. Indeed, there are clear examples of multi-phase outflows in GPS/CSS, seen in neutral hydrogen or warm ionized gas using the [O\,{\sc iii}]~$\lambda$5007 emission line \citep{Holt08,Morganti18,Santoro18,Liao20,Mullaney13}. The detection rate for warm outflows seems to be higher in objects with outflows in H\,{\sc i}. The evidence points out that these outflows are driven by the radio source rather than by radiation pressure. Moreover, results gathered by \cite{Mullaney13} support earlier findings that compact radio sources display more extreme [O\,{\sc iii}]~$\lambda$5007 kinematics than larger radio sources \citep{Holt08} and indicate that compact radio sources do indeed drive outflows. From the molecular gas perspective, \cite{Morganti22} found molecular outflows caused by the jet in the circumnuclear region, and also molecular gas around the jet in the region outside the host galaxy, forming a bubble that heats the pre-existing gas. This is consistent with the maintenance mode scenario, which prevents the gas from cooling. 

Despite the wealth of information already available on GPS/CSS, several open questions are still the topic of discussion. For instance, \textit{what is the proportion of compact sources with nuclear outflows? Is it similar to that of extended radio sources? Is the outflowing gas kinetic energy in compact galaxies as powerful as that found in extended radio galaxies?} While numerical simulations and estimates of jet power suggest that compact radio sources can drive outflows out to size scales of several times the radio source size \citep{Zovaro19}, the lack of suitable data precludes the comparison of theory with observations.

To contribute to the answer to the above questions, in this work, we study a sizeable sample of compact sources selected from the work of \cite{Nascimento22}, who analyzed 82 CSS/GPS using SDSS optical spectra to study the stellar population of these sources. Our goal is to carry out the first systematic study of the properties of the ionized gas in such objects in the optical range and establish the influence of the compact radio jet in the ISM through the study of the properties of the ionized gas.

The parent and control samples are described in Section~\ref{sec:observations}. The analysis of the integrated spectra along with the calculation of the physical parameters related to the outflows is presented in Section~\ref{sec:integratedspectrum}. We discuss the results in Section~\ref{sec:discussion} and our conclusions are in Section~\ref{sec:conclusions}. Standard cosmological parameters are employed in this paper: $H_0 = 67.8$ km\,s$^{-1}$Mpc$^{-1}$ and $\Omega_m$ = 0.308.

\section{Sample description }
\label{sec:observations}

%data

\begin{figure*}
\includegraphics[width=\textwidth]{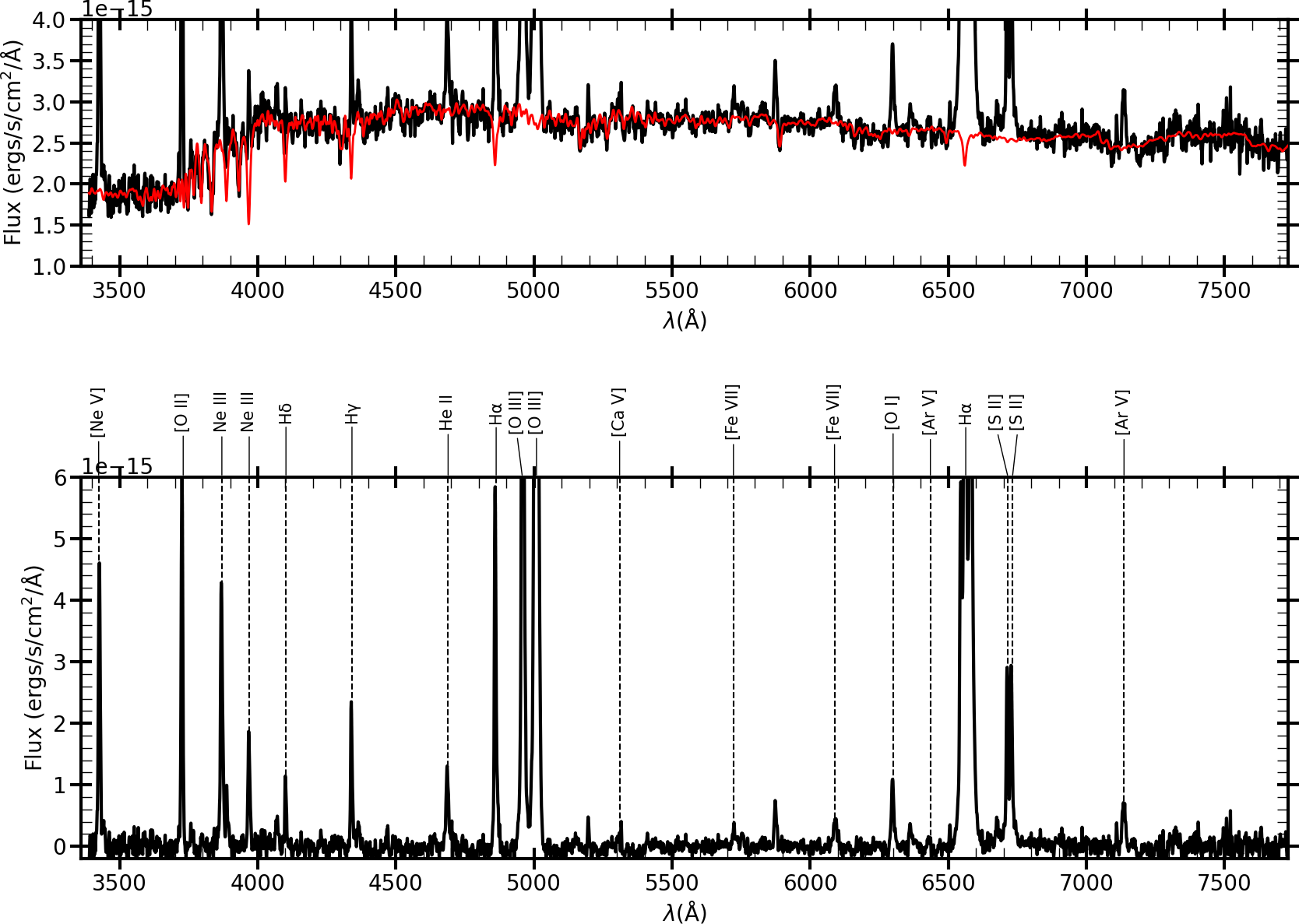}
\caption{In the top panel, we plot the spectrum of J0945+1737 (in black) before subtracting the stellar population. The stellar population template adjusted by {\sc starlight} is in red. The lower panel shows the spectrum after subtraction of the stellar population along with the identification of the most prominent emission lines. } \label{fig:sub-stellar-pop}
\end{figure*}

Our data were selected from the sample of \cite{Nascimento22}, which was already pre-selected from publicly available radio catalogues  \citep{Spencer89,Fanti90,Fanti01,Stanghellini98,Peck00,Kunert02,Snellen98,Snellen04,Tengstrand09,Hancock10,Bajraszewska10,Son12,Jeyakumar16,Jarvis19}, containing the objects with z~$<1$. From this dataset, objects with optical counterparts from the Sloan Digital Sky Survey (SDSS) \citep{Alam15} were further identified, resulting in 84 CSS/GPS. \cite{Nascimento22} also included SDSS spectra of 22 MPS classified as such by the GaLactic and Extragalactic All-sky Murchison Widefield Array Survey \citep[GLEAM,][]{Wayth15,Hurley16}. From this parent sample of 106 AGN, we selected spectra where the emission lines, in particular, \oiii{}~$\lambda5007$ are observed at $>3\sigma$ relative to the continuum. Among the selected sources, two presented linear sizes above 20~kpc and are classified as FRII radio galaxies. These two sources were removed from our sample. Thus, our final sample comprises 73 CSS/GPS and 9 MPS. The selected objects are listed in Section \ref{sec:appendixA}. In the remainder of this paper, we refer to these 82 CSS, GPS, and MPS as compact sources for simplification. In all the spectra studied here previously to any analysis we removed the AGN and stellar continuum as fitted by \cite{Nascimento22}. In Figure \ref{fig:sub-stellar-pop} we present an example of the spectrum before (upper panel) and after (lower panel) subtracting the stellar population (in red) for the galaxy J0945+1737.

While \cite{Nascimento22} studied these galaxies from the perspective of the stellar population, here we analyze the emission lines of high, medium, and low ionization and their relationship with outflows. Moreover, we characterize the compact sources based on the black hole mass, the mass of the outflowing gas, gas kinematics, and physical parameters such as the outflow rate, the kinetic energy, and the kinetic power. We assume that the detection of a broad component in \oiii{}~$\lambda5007$ is a signature of the presence of a nuclear outflow. Our analysis will allow us to compare the kinetic power of the gas with the AGN bolometric and radio luminosities to determine if the outflow can efficiently remove the hot gas from the host galaxy.

\subsection{The control sample}
\label{sec:control-sample}

To compare the results obtained for the compact galaxies with those previously found for extended radio-loud AGN, we use a subsample of the results derived by \cite{Speranza21} (hereafter S21). Their sample is composed of 37 galaxies with redshift $<0.3$ from the 3C~Catalogue, divided into three classes based on the optical emission lines; high-excitation galaxies (HEGs), low-excitation galaxies (LEGs), and broad-line objects (BLO). We notice that the latter group belongs to the high-excitation class with broad lines. In their analysis of the \oiii{}~$\lambda5007$ emission, they found 19 objects with nuclear outflows, obtained after integration of the innermost $0.6"\times0.6"$ region. They calculated outflow parameters and studied them for the high and low-excitation subclasses. These 19 RG with nuclear outflow compose our control sample. Some of the methods used here are similar to those employed by S21 to carry out a fair comparison of compact galaxies and FRI/FRII (see Section~\ref{sec:integratedspectrum}).

\section{Analysis of the integrated spectra}
\label{sec:integratedspectrum}

To study the physical properties of the emission gas and the outflow properties of the compact sources we characterize the observed emission line profiles in terms of their integrated flux, the peak centroid, and the full width at half maximum (FWHM). 
To this purpose, we fit the line profiles with the free available open-source tool {\sc pyspeckit}  \citep{Ginsburg11,Ginsburg22}. It is a Python-based package for spectral analysis that allows the measurement of individual lines or sets of blended lines. We found that spectral features in the compact sample like H$\beta$, \oiii{}~$\lambda\lambda4959,5007$ and H$\alpha$ were modeled with up to three Gaussian components. Other lines such as \rm{[N \sc{ii}]}~$\lambda\lambda6548,6584$ and \rm{[S \sc{ii}]} $\lambda\lambda6716,6732$ were satisfactorily reproduced with a single Gaussian component for every line. In this process, residuals after the fitting were less than 10\% of the line intensity.  It is important to notice that the FWHM measurements were corrected in quadrature by the SDSS instrumental resolution of $FWHM_{inst} = 3$ \AA. The fluxes and FWHM of the fitted emission lines are given in Section \ref{sec:appendixA}.

After the measurements of the observed fluxes, we corrected them for the effects of Galactic and internal extinction. To this aim, we employed the H$\alpha$/H$\beta$ ratio assuming an intrinsic value of 3.1 and the extinction law of \cite{Cardelli89}.  The ratio R between the extinction at V, A$_{\rm V}$ and the color excess between the B and V bands, E(B-V), was set to  R = 3.1. We notice that this correction was possible for a subsample of 47 out of the 82 galaxies, where both the H$\alpha$ and H$\beta$ lines were present within the spectral interval of SDSS or because both lines were 3$\sigma$ above the continuum. Hereafter, we will refer to these 47 objects as SCRG (subsample of compact radio galaxies). This subsample will be employed for most of the analysis done in the remainder of this paper unless otherwise stated.

\begin{figure}
\centering
\includegraphics[width=\columnwidth]{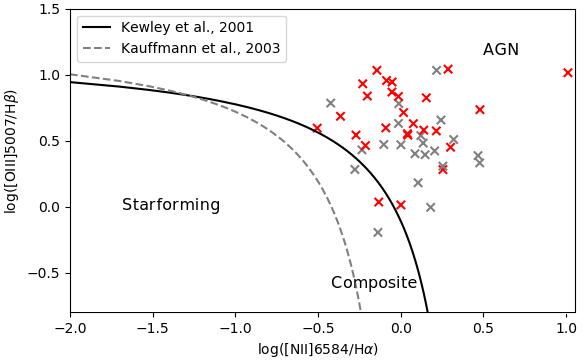}% 
\caption{BPT diagram of\oiii{}/H$\beta$ vs \rm{[N \sc{ii}]}/H$\alpha$ of the SCRG. The theoretical and empirical lines of \citet{Kewley01} and \citet{Kauffmann03} are the solid and dashed lines, respectively. The red crosses are the SCRG objects with outflowing gas signatures in their spectra. } \label{fig:bpt-outflows}
\end{figure}

With the extinction corrected fluxes of the emission lines, we examine the ionization state of the gas using the BPT diagram \oiii{}/H$\beta$ and \rm{[N \sc{ii}]}/H$\alpha$ \citep{Baldwin81}. The results are shown in Figure~\ref{fig:bpt-outflows}. The solid line represents the maximum starburst line from \citet{Kewley01}. The dashed line separates pure star-forming (SF) galaxies from composite galaxies by \citet{Kauffmann03}. If a galaxy presents a broad component (either because of a broad-line region or an outflow component) in any of the lines involved, only the flux of the narrowest component was employed for deriving its position in Figure~\ref{fig:bpt-outflows}. The points marked in red identify the sources where the \oiii{} line displayed a broad component (which we interpret as evidence of an outflow, see Sec.~\ref{sec:outflows}). 

It is easy to see that most galaxies in Figure~\ref{fig:bpt-outflows} lie in the locus of points occupied by AGNs. Only three sources are located in the transition region, indicating that they are composite objects, with AGN and star-forming characteristics. None of the targets are star-forming (H\,{\sc ii}) dominated sources.  We conclude that the SCRG analyzed here are AGN-dominated objects. That is, the observed emission line spectrum is AGN-driven.

\subsection{Emission line profiles}

\begin{figure}
\centering
\includegraphics[width=\columnwidth]{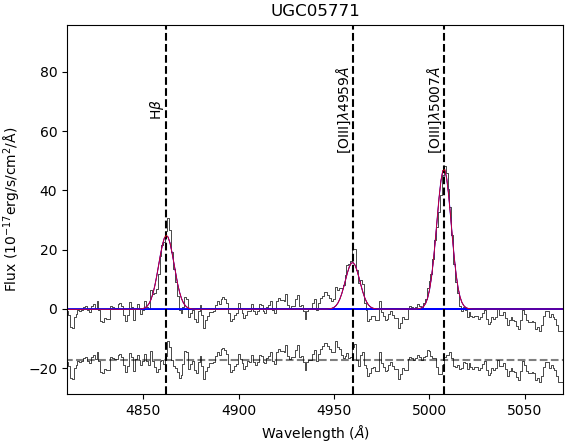}
\caption{Example of the emission line fitting of the observed \oiii{}~$\lambda\lambda$4959,5007 + H$\beta$ lines (in black) in UGC\,05771 with one Gaussian component (in red). The bottom panel shows the residuals after subtraction of the best-fit model to the observed profiles.} 
\label{fig:singlegaussian}
\end{figure}

\begin{figure}
    \centering
    \includegraphics[width=\columnwidth]{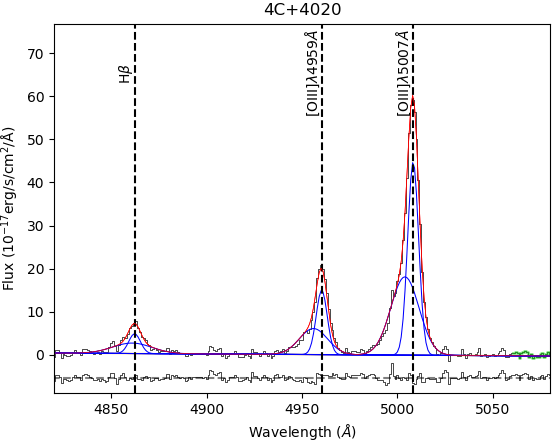}
    \caption{Same as Figure~\ref{fig:singlegaussian} but for 4C+40.20. In this example, the best solution was found after fitting the observed profiles with a narrow and broad blueshifted Gaussian component.}
    \label{fig:doublegaussian}
\end{figure}

\begin{figure}
    \centering
    \includegraphics[width=\columnwidth]{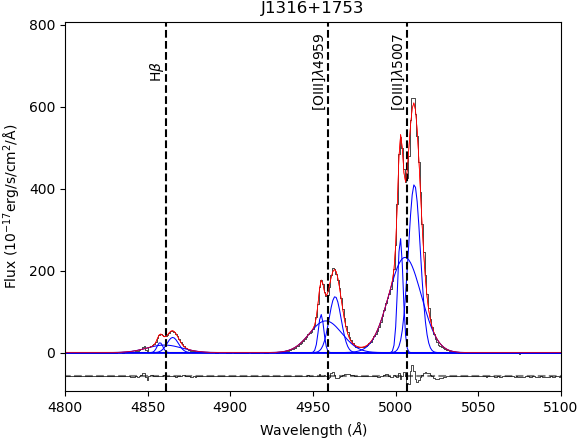}
    \caption{Same as Figure~\ref{fig:singlegaussian} but now for J1316+1753. Here, the lines were fit with a split narrow profile plus a broad component.}
    \label{fig:doublepeak}
\end{figure}

The line fitting procedure reveals a variety of emission line profiles based on the number and width of the Gaussian components, encompassing the presence of narrow, broad, and split emission line profiles. Figures~\ref{fig:singlegaussian} to~\ref{fig:doublepeak} display examples of the kind of \oiii{}~$\lambda5007$ profiles most commonly found. The fitting to the reminder of the sample is presented in Section~\ref{sec:apA:fitting}. They can be grouped into three main categories: the emission line is described by a single, narrow component (see Figure~\ref{fig:singlegaussian}); the emission line displays a narrow component plus a prominent broad component whose peak centroid can be either blueshifted or redshifted from the systemic velocity (Figure~\ref{fig:doublegaussian}); the emission line displays a narrow split line profile plus a broad blueshifted or redshifted component (Figure~\ref{fig:doublepeak}).

Regarding the sources with a single narrow component,  47 out of the 82 compact sources ($\sim58\%$) display such a characteristic. The FWHM of the \oiii{} line varies between 300~km\,s$^{-1}$ and 812~km\,s$^{-1}$, typical of what is found in classical AGN. As no evidence of asymmetries in the profiles is observed, the presence of a nuclear outflow of ionized gas cannot be confirmed.  

In 27 objects ($\sim$33\% of the compact sample) a single narrow plus a broad component in the \oiii{} lines are detected. We interpret the latter component as evidence of outflowing nuclear gas while the former is attributed to the contribution of the classical NLR. The example shown in Figure~\ref{fig:doublegaussian} clearly illustrates that scenario. The range of values of the FWHM for the narrower component is 300 - 764~km\,s$^{-1}$ while for the broader component, the FWHM varies between 430 and 2780~km\,s$^{-1}$. We also observed that in most cases the centroid position of the broad component is offset from the systemic velocity by values reaching up to 890~km\,s$^{-1}$.

We notice that among the galaxies with an outflow component, eight of them (10\% of the compact galaxies sample) present a split profile in the hydrogen and the forbidden lines. In four of them, the separation of the peaks is very clear, such as J1316+1753 (see Figure~\ref{fig:doublepeak}). In that source, three Gaussians were fitted in the \oiii{}~$\lambda5007$ forbidden line: two narrow components, of 300 and 500~km\,s$^{-1}$ to account for the blue and red split lines and one broad component, with a FWHM of $\sim$$1273$ km\,s$^{-1}$. Among the split-line category, we found broad components with FWHM varying from 1066~km\,s$^{-1}$ up to $\sim$2800~km\,s$^{-1}$.

Split \oiii{}~$\lambda5007$ optical emission has been detected previously in the literature and led \cite{Gerke07} to the idea that a binary AGN with two narrow line regions (NLR) produces two velocity components in the \oiii{} lines. In that way, the split forbidden emission lines could be the result of a binary system with two NLRs. Other explanations for the split lines were proposed (\cite{Fu12} for a review). This characteristic was found in optical spectra in several sources (such as \cite{Fu23,Speranza24}, for recent studies). Recent hydrodynamical simulations have shown that the expansion of the cocoon caused by the jet can result in a bipolar outflow emission symmetric to the galaxy center \citep{Mukherjee18b}.

\begin{figure}
\includegraphics[width=\linewidth]{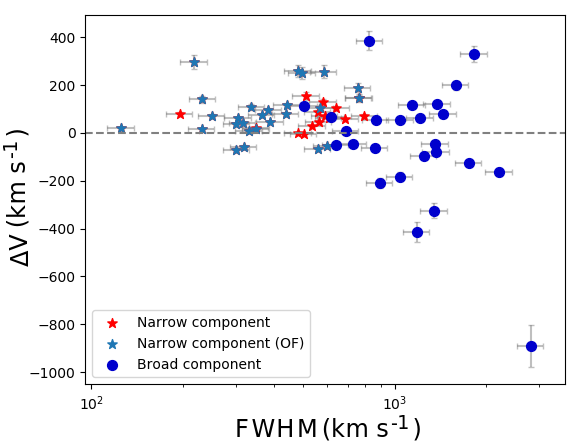}
\caption{Shift of the line center from the systemic velocity versus FWHM of each component fitted in the \oiii{}~$\lambda$5007 emission line. The dark blue points are the values of broad Gaussian components, the light blue stars are the values from the narrow component of sources with outflow (OF) and the red stars are the narrow component of sources without outflow signature.} \label{fig:deltav-fwhm}
\end{figure}

The distribution of the shift of the emission line center versus the FWHM of the Gaussian components fitted in the \oiii{}~$\lambda5007$ is shown in Figure~\ref{fig:deltav-fwhm}. The dark blue points are the values of broad Gaussian components fitted in objects with the presence of an outflow. The light blue and red stars are the values for the narrow components of the sources with and without the presence of outflows, respectively. The spread in the line shift is higher in broad Gaussian components (the blue dots in Figure~\ref{fig:deltav-fwhm}), reaching 1200~km\,s$^{-1}$, than in the narrow components (the blue and red stars). 
This justifies our initial assumption of the broad component representing outflowing gas. In fact, 4C12.50, widely known for displaying a prominent outflow, is located in the bottom right corner of the diagram.

\subsection{High ionized gas}

We investigate the presence of coronal lines (CLs) in the sample spectra. These lines are clear signatures of a strong X-ray continuum typical of AGN because of the ionization energies (IP$>$~100 eV, where IP is the ionization potential) needed to form the ion emitting the CL. For this reason, that emission is free of any contribution of star-forming regions and is regarded as an unambiguous indicator of the AGN activity in the ionized gas \citep{Ardila06}.

In radio-loud sources, evidence of CLs was previously reported by, among others, \citet{tadhunter87} and \citet{tadhunter88}. They detected \rm{[Fe \sc{vii}]} emission in the radio-galaxy PKS\,2152-69 at $\sim$~8~kpc from the galaxy's nucleus. That emission was found co-spatial with the radio jet, suggesting that the coronal emission is produced by the passage of the radio jet and its interaction with the ISM. \citet{speranza22} reported the detection of extended CL employing the \rm{[Si \sc{vi}]}~19631\AA\ in the Teacup galaxy, also classified as a radio-loud AGN. Moreover, examples of the presence of extended CL emission associated with the radio-jet in radio-weak AGN abound in the literature \citep{Ardila06,May18,Faria21,faria23}. Therefore, it is expected the presence of CL emission in the sample of compact radio sources.

We search for the presence of CLs, in particular for [Fe\,\sc{vii}]~6087\AA, \rm{[Fe \sc{x}]}~6374\AA, and \rm{[Ne \sc{v}]}~3425\AA, the strongest CLs in the optical region, to find additional evidence of jet-driven outflows. \citet{Ardila06}
had already presented convincing evidence of the relationship between CLs and shocks likely produced by the interaction of the jet with the ISM. That result was supported by theoretical models that combine the effects of shocks and radiation from the AGN.

If we consider only the SCRG sources in the BPT diagram, 10 are coronal line emitters (22\%). In the compact galaxies sample AGNs, 18 sources display at least one CL in their spectra. Table~\ref{tab:emissao-coronal} lists these 18 CL emitters along with the emission line fluxes measured (columns 2-3). We also calculated the emission line ratios \rm{[Ne \sc{v}]}~3425\AA/H$\alpha$ (column 4) and \rm{[Fe \sc{vii}]}$~\lambda6087$/\oiii{}~$\lambda5007$ (column 5). It can be seen that  \rm{[Ne \sc{v}]}~$\lambda3425$ (IP = 97.11 eV) is the most common CL in the sample, being detected in 15 sources. We notice that in the three AGNs where it was not measured, the spectra did not cover the spectral region where that line is located. \rm{[Fe \sc{vii}]}$~\lambda6087$ was found in 11 sources while \rm{[Fe \sc{x}]}~$\lambda$6374 was not detected in any of the objects of the sample. Finally, 8 AGNs have both \rm{[Ne \sc{v}]} and \rm{[Fe \sc{vii}]}.   

\begin{figure}
\includegraphics[width=\linewidth]{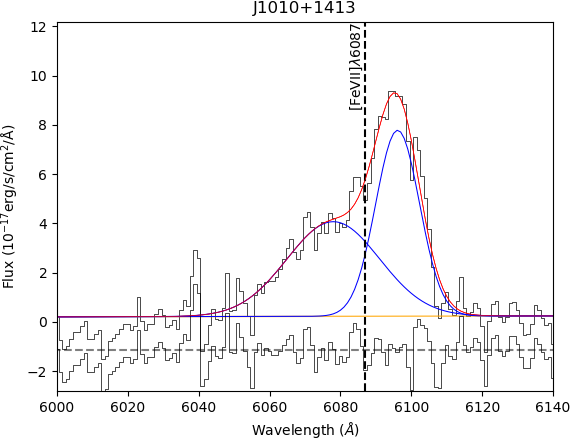}
\caption{Example of the blue-shifted, broad component fitted to the line profile in \rm{[Fe \sc{vii}]}~$\lambda$6087 in J1010+1413. The black line is the observation, the blue lines are the two Gaussian components and the red line is the fit. } \label{fig:fe7-dp}
\end{figure}

In terms of the emission line profile, 7 out of 11 \rm{[Fe \sc{vii}]} sources present a broad emission in addition to the narrow component, as seen in the example shown in Figure~\ref{fig:fe7-dp}. It is interesting to note that all of these sources also present outflow components in \oiii{} lines. The other four sources have low signal-to-noise (S/N) and there was no detection of other components.

To compare the detection of CL in compact sources relative to that in extended radio galaxies, we investigate the presence of CL in the HEGs RG sample of S21, which are all FRII objects. We found at least one CL in 10 out of 18 objects ($55\%$).  
Thus, the coronal emission is more frequently found in extended RG than in our sample of compact objects, where just $\sim$22\% displays such features.  

The 22\% of CL frequency found in compact galaxies sample is significantly smaller than the frequency of radio weak AGN \citep[where the percentages vary from 45\% to 67\%]{denBrok22,Ardila11}. In the context of radio galaxy evolution, we interpret the lower frequency of CLs in the compact sample because it does not distinguish clearly among future FRI or FRII sources, assuming that a fraction of the compact sources will evolve to either one of the RG classes. Moreover, we notice that the CL frequency was carried out in the whole sample, where a fraction of the sources does not show evidence of AGN activity based on the optical spectrum. Therefore, it is very likely that these latter sources would not be located in the AGN locus of points in the BPT Diagram of Figure~\ref{fig:bpt-outflows}.   

It is also important to highlight that some of the spectra of the compact objects have a lower signal-to-noise ratio than the extended RG. That may also help to explain the observed differences in CL frequency.

Regarding the ratio between \rm{[Fe \sc{vii}]} and \oiii{} (last column of Table~\ref{tab:emissao-coronal}), the median value calculated for the compact galaxies is 0.04. This is similar to the values obtained in the literature for radio-weak AGN, in which \rm{[Fe \sc{vii}]}/\oiii{} ranges from 0.01 up to 0.08 in the nuclear region \citep{Mazzalay10}. From this comparison, we see that there is no indication of an influence of the compact jets in the high excitation emission of the gas. However, it is important to notice that any possible shock-driven extended CL region at scales of hundreds of parsecs will be washed out by the SDSS aperture, which should be dominated by the much brighter, nuclear component.

\begin{table*}
    \centering
    \caption{Integrated fluxes of the [Ne \sc{v}]~$\lambda3425$ and \rm{[Fe \sc{vii}]}~$\lambda6087$ emission lines and line ratios \footnote{The integrated fluxes in the first two columns are not corrected by reddening. The line ratio in the two last columns is between the corrected fluxes of the emission lines.}.}
    \label{tab:emissao-coronal}
    \begin{tabular}{lllll}
    \hline
     Source      & Flux [Ne \sc{v}]$^{**}$ &  Flux \rm{[Fe \sc{vii}]}$^{**}$ & [Ne \sc{v}]/H$\alpha$ & \rm{[Fe \sc{vii}]}/\oiii{}  \\
             & ($10^{-15}$ erg\,s$^{-1}$) &  ($10^{-15}$ erg\,s$^{-1}$)  &   &    \\
    \hline 
    \hline
4C+47.27          &  0.43 $\pm$ 0.05   &  --         &  --     &  --     \\
4C+40.20          &   0.26 $\pm$ 0.05   &  --        &  --     &  --     \\
SDSSJ09260        &  0.29  $\pm$ 0.04   &   --       &   --    &    --   \\
J0945+1737       &  6.09  $\pm$ 0.05  &   1.18  $\pm$ 0.08     &  0.54    &  0.02     \\
J1000+1242       &  3.37  $\pm$ 0.06  &   0.94 $\pm$ 0.10       &  0.35     &   0.04    \\
J1010+1413$^{*}$   &   3.44 $\pm$ 0.11  &   1.14 $\pm$ 0.25      &   0.27    & 0.01      \\
J1100+0846$^{*}$        &   --    &    0.58 $\pm$ 0.11  &  --     &    0.04   \\
4C+39.32          &   0.05 $\pm$  0.05  &    --      &  0.05     &    --   \\
{[}HB89{]}1153    &   0.91  $\pm$  0.07  &    --      &  --     &  --     \\
B3-1154+435      &   1.24  $\pm$ 0.04    &  1.23 $\pm$ 0.04        &  0.02     &  --     \\
SBS1250+568      &  1.63   $\pm$  0.05   &  4.07 $\pm$ 0.06    &  0.01     &  --     \\
J1316+1753$^{*}$  &   3.59 $\pm$ 0.07 &    0.20 $\pm$ 0.11     &   --    &   --    \\
3C268.3           &   0.15 $\pm$ 0.05  &   --        &   --    &   --    \\
J1338+1503       &  3.14  $\pm$  0.06 &    0.60  $\pm$ 0.09    &  0.73     & 0.05      \\
J1356+1026       &  7.65 $\pm$  0.07 &   0.83  $\pm$ 0.12     &   --    &   --    \\
NGC5506$^{*}$     &  --        &  0.88   $\pm$ 0.08     &  --     &  0.005      \\
2MASXj15          &  --    &  4.28    $\pm$ 0.14     &    --   &   0.40    \\
PKS1607+26        &   0.08 $\pm$  0.05   &  --        &   --    &    --   \\
    \hline     
    \multicolumn{5}{c}{(*) Two Gaussian components fitted in the emission line.} \\
    \multicolumn{5}{c}{(**) Fluxes before reddening correction.}
    \end{tabular}
\end{table*}

\subsection{Black hole mass estimation}
\label{sec:bhmass-estimation}

To study the accretion mode of the AGN and also to characterize our sample in terms of black hole mass, we computed these quantities for the 82 compact galaxies. As described in Section~\ref{sec:integratedspectrum}, 58\% of our sample do not show broad lines in their spectra, which is characteristic of AGN with central emission obscured by the torus. However, the sample does show stellar absorption lines. Thus, to obtain the BH masses for the entire sample regardless of the presence of broad emission lines in the spectra, we employed the M-$\sigma_*$ relation described by \cite{Kormendy13}:

\begin{equation}
    \log \left( \frac{M_{\rm BH}}{M_{\odot}} \right) = 8.49 + 4.38 \times \log \left( \frac{\sigma_*}{200\,{\rm km\,s^{-1}}} \right)
\end{equation}

\begin{figure}
\includegraphics[width=\columnwidth]{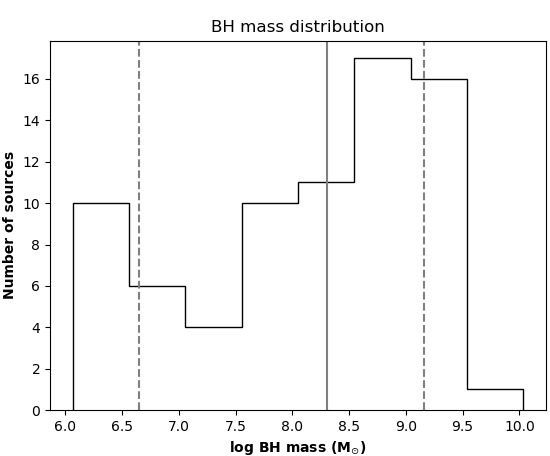}
\caption{Black hole mass distribution of the compact sources. The vertical grey solid line shows the median value (8.3) of the distribution, and the dashed lines limit the range between the 16th and the 84th percentiles.} \label{fig:bhm-histogram}
\end{figure}

The stellar velocity dispersion $\sigma_*$ was previously obtained through the fitting of the stellar contribution in the spectra by \cite{Nascimento22}. The results of black hole masses and the respective values of velocity dispersion for each AGN are given in Section~\ref{sec:appendixA}. We found that the BH masses range between $10^{6}$M$_{\odot}$ and $10^{10}$M$_{\odot}$, that is a interval similar to previous results of \cite{Liao20} for compact sources. We notice that both samples have 60 sources in common. \citet{Liao20}  had shown that there is a superposition in the black hole mass range between compact and extended sources, except for some compact galaxies with BH masses lower than $\sim$$10^8$~M$_{\odot}$. Sources below that threshold are not expected to evolve to extended RG, favoring the frustration scenario in the evolution of compact sources. The distribution of black hole mass of our sample is shown in Figure~\ref{fig:bhm-histogram}, where 60\% of the compact sources are located in the range $\sim$$10^{8}$M$_{\odot}$ to $10^{10}$~M$_{\odot}$. The determination of the BH masses will allow us to establish the presence of possible correlations between that parameter and those related to the outflow, such as the outflow mass, the outflow rate, and the kinetic power. This is described in Section~\ref{sec:outflows}. 

With the BH mass, the black hole accretion rate can be estimated using the Eddington ratio $R_{\rm Edd}$:

\begin{equation}
\label{eq:edd-ratio}
    R_{\rm Edd} = \frac{L_{\rm bol}}{L_{\rm Edd}}
\end{equation}

\noindent where $L_{bol}$ is the bolometric luminosity and $L_{\rm Edd}$ is the Eddington luminosity. The former is calculated using the luminosity of the narrow component of  \rm{[O \sc{iii}]}~$\lambda5007$ multiplied by a factor of 600 (see Section~\ref{sec:outflows}). The latter is obtained using the expression $L_{\rm Edd} = 1.38 \times 10^{38}~M_{\rm BH}/M_{\odot}$~erg\,s$^{-1}$. We found values of Eddington ratio $R_{\rm Edd}$ varying from $10^{-3.1}$ to $10^{0.7}$. From these, nine sources have $R_{Edd}>0.01$, which indicates that these sources have a radiatively efficient accretion disk. 
In the \cite{Liao20} sample, the Eddington ratio of CSS sources ranges from $\sim10^{-4.93}$ to $\sim10^{-0.37}$. We attribute this difference to the sample composition, and the procedures used to obtain the Eddington ratio in their work, which differs from ours. First, they used emission lines (H$\beta$, \rm{[O\sc{iii}]}~$\lambda5007$ or H$\alpha$) to derive BH masses, while we used the M-$\sigma_*$ relation. Also, they did not mention performing a reddening correction of the emission line fluxes, which underestimates the line fluxes. This difference will propagate and affect the BH mass (which mean value is $10^{8.64}$M$_{\odot}$ for young RG, higher than our of $10^{8.11}$M$_{\odot}$), and giving smaller values of $R_{Edd}$.

\subsection{Outflowing gas parameters}
\label{sec:outflows}

Our aim here is to determine the frequency to which nuclear outflows are detected in compact galaxies and, if an outflow is present, to characterize it by its physical properties.

As presented in Section~\ref{sec:integratedspectrum}, nearly half of the 82 compact sources present broad forbidden emission lines in their spectra ($\sim$43\%). The presence of these broad emission lines is attributed to outflowing gas in the inner regions of the AGN (in sub-kiloparsec scales) and may provide important insights into the ionization mechanisms and the properties of the ejected gas. Here we investigate further the role of the jet in the ISM by calculating the physical properties of the outflow in the compact sources. Moreover, we compare these properties to optical and radio parameters of the sample as well as in a control sample of extended RG (Section~\ref{sec:control-sample}) to place the compact sources in the whole RG context.  

Among all galaxies with detected outflow components, outflow parameters could only be derived for 19 sources (listed in Table~\ref{tab:of-prop}). This limitation arises because only these sources exhibit both H$\alpha$ and H$\beta$ emission lines above 3~$\sigma$, necessary to calculate reddening correction of their fluxes. 

\subsubsection{Physical parameters}

We calculate the properties of the outflow of the compact sources by estimating their mass, mass rate, energy, and kinetic power. The mass of the outflowing gas can be obtained using the \oiii{}~$\lambda5007$ luminosity $L_{44}^{(\rm [OIII])}$ of the broad component using the equation by \cite{CanoDiaz12}:

\begin{equation}
\label{eq:outflowmass}
    M^{out}_{ion} = 5.3 \times 10^7 \frac{L_{44}^{\rm{[O III]}}}{n_{e3} 10^{[O/H]}} {\rm(M_{\odot})}
\end{equation}

 \noindent where the luminosity is given in units of $10^{44}$ erg\,s$^{-1}$, $n_{e3}$ is the electron density in units of $10^{3}$ cm$^{-3}$, and $10^{\rm [O/H]}$ is the oxygen abundance in Solar units. To this purpose we compute the electron density using the \rm{[S \sc{ii}]} emission line flux ratio \rm{[S \sc{ii}]}~$\lambda6716/6732$ ($R$) as in \cite{Proxauf14}.

We used only narrow components to derive the line ratio $R$. For the sources that have no \rm{[S \sc{ii}]}~$\lambda\lambda6716,6732$ lines in their spectra, the mean value of 300~cm$^{-3}$ was used. It is important to note that estimating the electron density using the sulfur lines \rm{[S \sc{ii}]}~$\lambda\lambda6716,6732$ may underestimate the actual density by up to two orders of magnitude \citep{Santoro20}. A more precise estimation can be done by using the transauroral lines \rm{[S \sc{ii}]}$\lambda4076,4068$, but these lines are barely detected in our data. Nevertheless, the density estimation by the red sulfur line ratio can give us an upper limit to the mass going out of the central region. 
 
\begin{figure}
\centering
\includegraphics[width=\columnwidth]{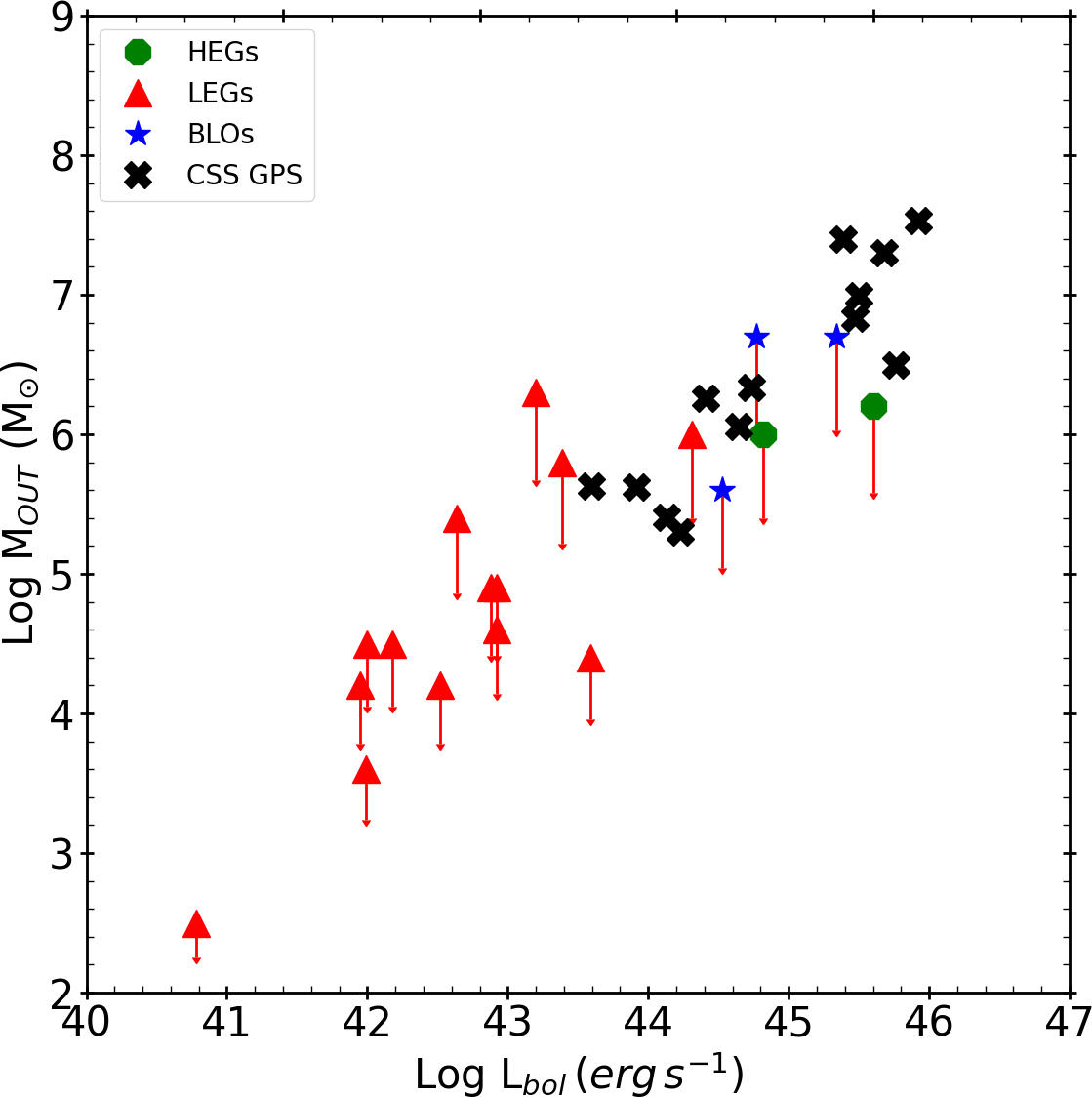}
\caption{Outflow mass versus bolometric luminosity for compact galaxies (black) of our sample and FR galaxies (colored) of S21. The latter sample data are divided into high-excitation, broad-line objects, and low-excitation (HEG, BLO, and LEG, respectively). }
\label{fig:outflowmass}
\end{figure}

With the calculated densities and assuming a Solar value for the oxygen abundance in equation~\ref{eq:outflowmass}, we estimated the outflow masses, $M_{of}$. The results are reported in the fourth column of Table~\ref{tab:of-prop}. The values of $M_{of}$ range between 2.0$\times$10$^{5}$~$M_{\odot}$ and 3.4$\times$10$^{7}$~$M_{\odot}$. To estimate the bolometric luminosity we used the relation $L_{\rm bol}=600\times L_{\rm [OIII]}$, where the \oiii{} luminosity refers to the luminosity of the narrow component of the $\lambda5007$ line.  The same procedure is made in S21 so that a fair comparison can be established with both sets of data. The bolometric luminosities are listed in the second column of Table~\ref{tab:of-prop}. We found values between $\sim$10$^{44}$erg\,s$^{-1}$ and 10$^{47}$erg\,s$^{-1}$. The relation between $M_{out}$ and $L_{bol}$ are plotted in Figure~\ref{fig:outflowmass}, where the black dots are our data, and the red, green, and blue ones are LEGs, HEGs, and BLOs of S21, respectively.  We see from Figure~\ref{fig:outflowmass} that the outflow mass increases with bolometric luminosity, and these values are comparable to the nuclear ones of high excitation extended radio galaxies of \cite{Speranza21} (colored points in Figure~\ref{fig:outflowmass}). This result shows that compact jets can drive amounts of gas out of the galaxy center comparable to that of jets in extended radio galaxies. Moreover, the bolometric luminosity of compact sources with outflow components is, on average, three orders of magnitude brighter than that of LEGs.

\begin{figure}
    \centering
    \includegraphics[width=\columnwidth]{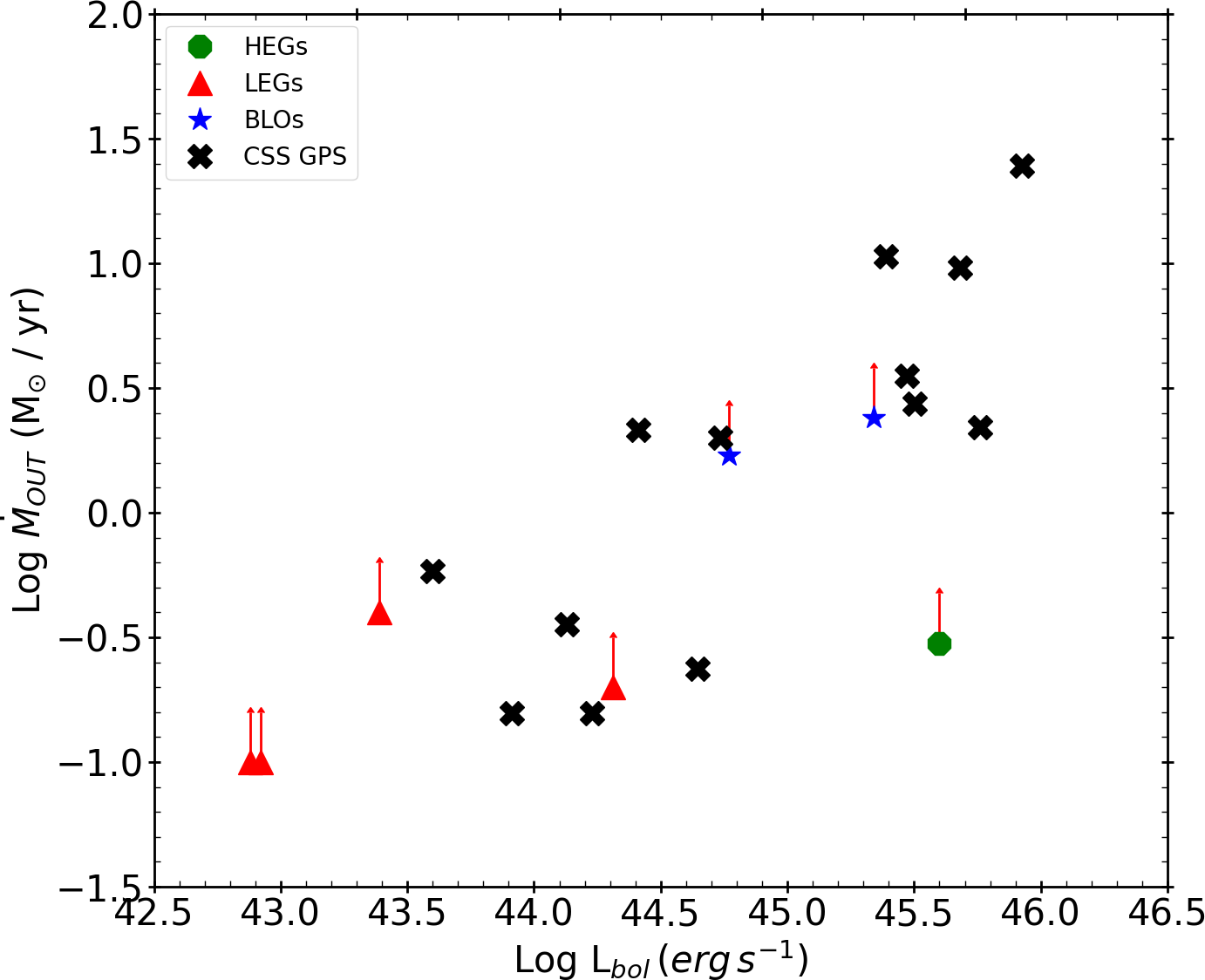}
    \caption{Outflow rate versus bolometric luminosity. The symbols are the same as Figure~\ref{fig:outflowmass}.}
    \label{fig:outflowrate}
\end{figure}

\begin{figure}
    \centering
    \includegraphics[width=\columnwidth]{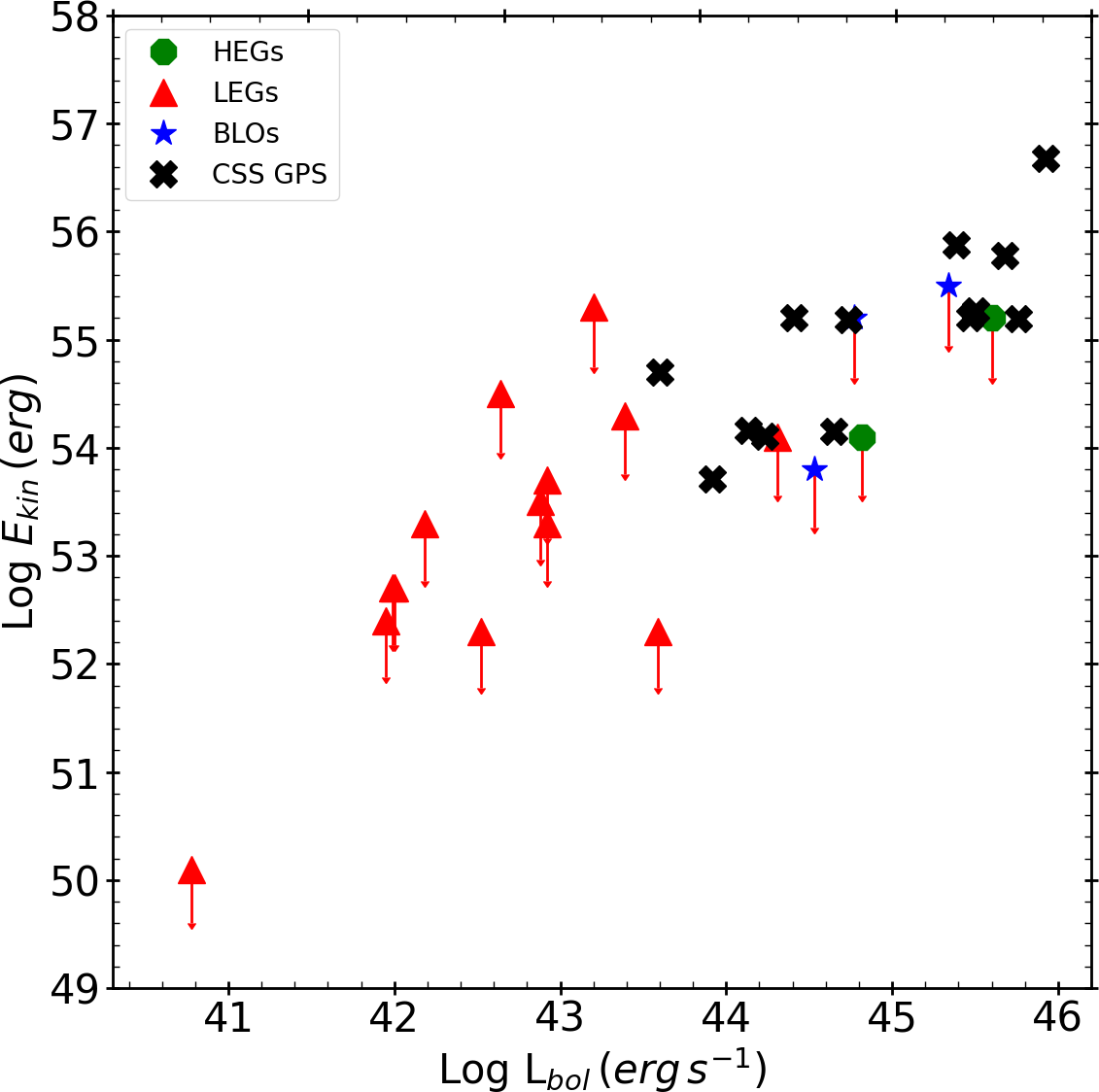}
    \caption{Kinetic energy versus bolometric luminosity. Black crosses are the compact sources of this work, and colored data are extended radio galaxies of G. Speranza (priv. communication).}
    \label{fig:kineticenergy}
\end{figure}

\begin{figure}
    \centering
    \includegraphics[width=\columnwidth]{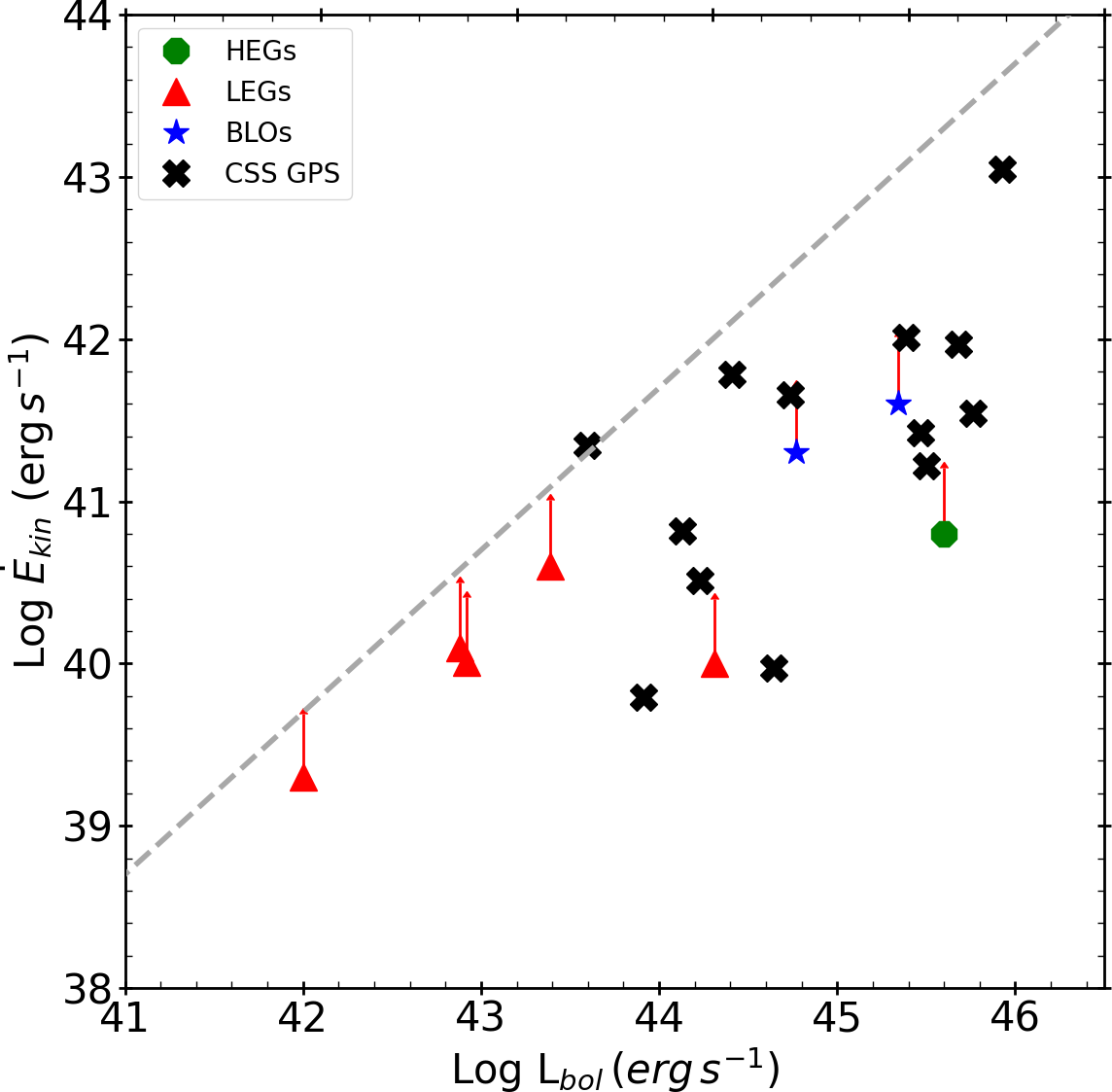}
    \caption{Kinetic power versus bolometric luminosity. The colors are the same as Figure~\ref{fig:outflowmass}.  The grey dashed line traces the $\dot{E}$ = 0.5\% $L_{\rm bol}$.}
    \label{fig:kineticpower}
\end{figure}

The outflow rate can be obtained using the \oiii{}~$\lambda5007$ luminosity of the broad component in the equation 

\begin{equation}
\label{eq:outflow-rate}
    \dot{M}^{out}_{ion} = 164  \frac{L_{44}^{\rm{[O III]}} v_3}{n_{e3} 10^{[O/H]} R_{out}} {\rm (M_{\odot} yr^{-1})}
\end{equation}

\noindent where $v_{3}$ is the outflow velocity in units of 1000 km\,s$^{-1}$ and $R_{out}$ is the radius of the outflowing gas. To compute the gas velocity we used the centroid position of the broad component of \oiii{}~$\lambda5007$ relative to the systemic velocity. Then we sum half of the value of the FWHM of the broad component. This approach gives us the median velocity of the outflow. A similar procedure is used in S21 in which they calculate their median velocity $v_{50}$, so we can compare the outflow rates of both samples. Another approach is to use the maximum velocity ($v_{max}$), which gives the maximum velocity reached by the outflow considering projection effects, and encompasses 95\% of the broad line flux. This procedure was made to obtain a range of outflow rates and this is discussed in Section~\ref{sec:discussion}. The outflow rates are plotted in Figure~\ref{fig:outflowrate}.

As our sources are compact in terms of their radio emission, we do not expect outflow radii higher than the seeing of the observations. In this scenario, we considered that the outflows are nuclear ones. Therefore, the radius used in equation~\ref{eq:outflow-rate} equals the 1.5 arcsec of SDSS aperture times the kpc\,arcsec$^{-1}$ scale of each source. This approach allows us to have individual values of outflow radius, instead of using a mean value for the entire sample. The radii are shown in column 3 of Table~\ref{tab:of-prop}. Also, it is important to note that this approach gives us outflow radius in kiloparsec scales, the same order as the ones obtained for extended RGs in S21, allowing us to compare directly the gas parameters of the compact and the extended RG.

To compute the kinetic energy and kinetic power of the outflowing gas we use the values of $M_{out}$ and $\dot{M}_{out}$ already derived and the equations shown by \cite{Holt06}:

\begin{equation}
    E = 1.98\times 10^{43}\frac{1}{2}M_{out} \nu^2 \ \ {\rm(erg)}
\end{equation}

\begin{equation}
\label{eq:kineticpower}
    \dot{E} = 6.34 \times 10^{35}\frac{1}{2}\dot{M}_{out} \nu^2 \ \ {\rm(erg\,s^{-1})}
\end{equation}

\noindent where $\nu$ is the velocity of the outflow gas, the same used in equation~\ref{eq:outflow-rate}. It is important to highlight that in the estimation of the kinetic power, there are uncertainties associated with the density estimation, as discussed before, and also in the radius of the outflow, once it is obtained by an estimation using the observed seeing. The values obtained from the evaluation of these parameters are therefore considered as upper limits for the outflow mass and rates and lower limits for the kinetic energy and power. 

The kinetic energy and kinetic power are shown in Figures~\ref{fig:kineticenergy} and \ref{fig:kineticpower}, respectively. We see that in the compact galaxies these quantities, as well as the outflow mass and outflow rate obtained before (see Figures~\ref{fig:outflowmass} and~\ref{fig:outflowrate}), are comparable to that of the control sample of extended radio sources. In Figure~\ref{fig:kineticenergy} we see that the kinetic energy of the outflow of our data (black crosses) and the HEGs and BLO nuclear outflows (green circles and violet stars, respectively) occupy the same locus of points. Furthermore, we see from Figure~\ref{fig:kineticpower} that only one compact source present kinetic power above the $\dot{E}$ = 0.5\% L$_{\rm bol}$ line (dashed grey). We also see that the kinetic power of our data is about two orders of magnitude lower than that of the nuclear outflows of extended RG, and up to two orders of magnitude above that of LERG in bolometric luminosity. 
When comparing the kinetic power with the black hole mass of the compact sources, we see no explicit trend.

\begin{figure}
\includegraphics[width=\linewidth]{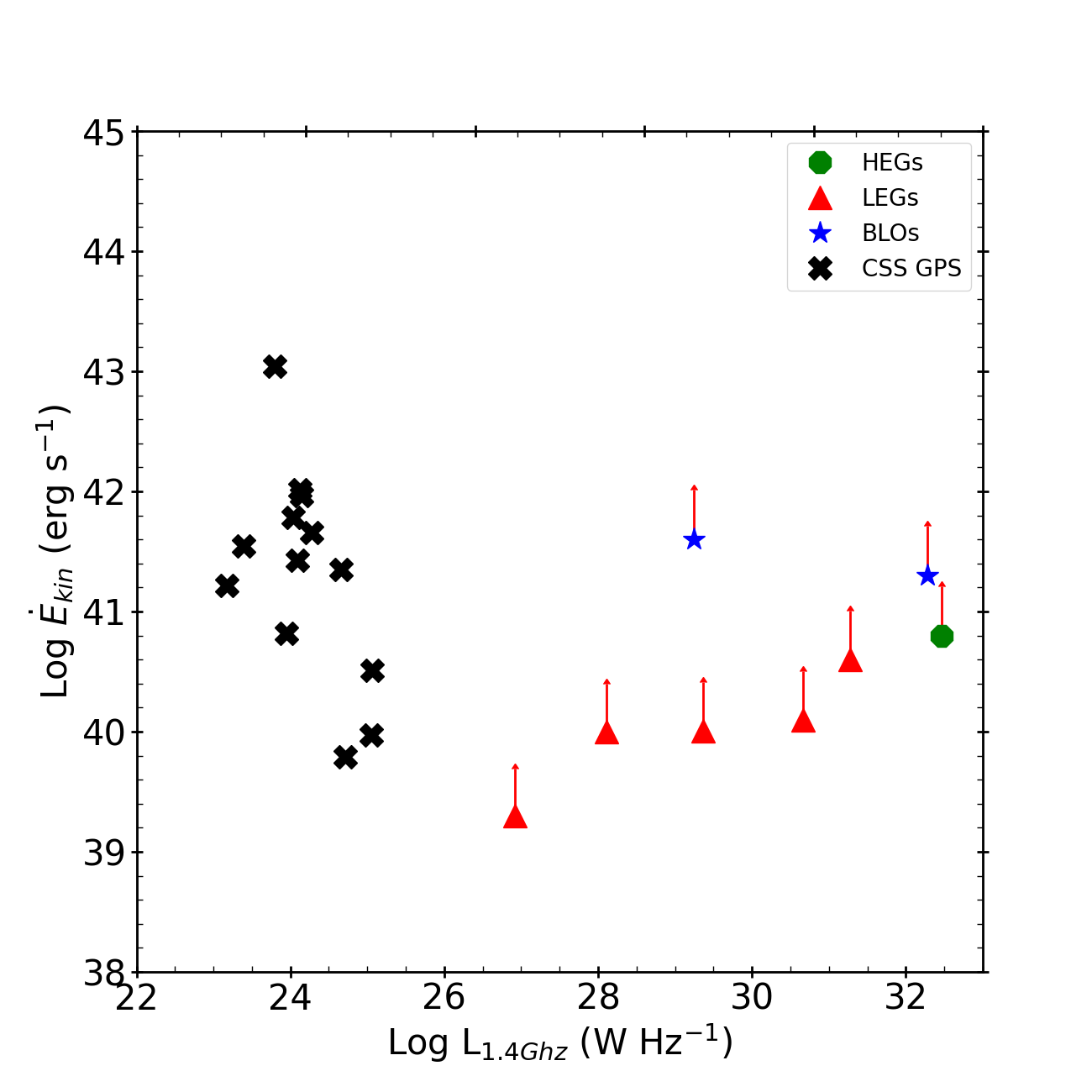}
\caption{Kinetic power versus radio luminosity in 1.4~GHz for the compact sources of this work (black crosses) in comparison with extended RG by S21 (red triangles, green circles, and violet stars).}
\label{fig:kineticpowerxradio}
\end{figure}

In Figure~\ref{fig:kineticpowerxradio} we show the kinetic power as a function of the radio luminosity for the compact sources of this work and the RG of the S21 sample. Despite compact and extended sources displaying similar values of kinetic power, their radio luminosity clearly separates compact and extended radio sources.

\section{Discussion}
\label{sec:discussion}

\begin{figure}
\includegraphics[width=\columnwidth]{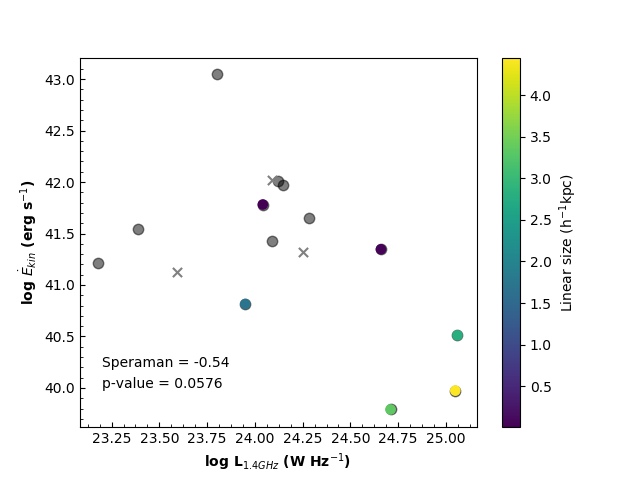}
\caption{Kinetic power versus radio luminosity for compact galaxies with linear size (LS) scale. Grey dots have no data for LS. The grey crosses are values obtained by \cite{Speranza24}. The Spearman correlation coefficient is -0.54 and the p-value is 0.058.} 
\label{fig:bestfit-kinpower}
\end{figure}

In this work, we analyze the frequency of outflows in the spectra of compact radio galaxies, as well as the physical properties of the outflowing gas detected using optical spectroscopy. We consider a broad component in \oiii{}~$\lambda5007$ as a signature of the outflow. We found this signature in almost half of our sample (43\%).

To investigate how common are the outflows in radio galaxies, we compare our results to the ones obtained previously for radio-weak sources in the literature. In the work of \cite{Singha22} with a small sample of local radio-weak sources (36 AGN),  they reported \oiii{} wings as signatures of ionized outflows in all objects of the sample (the CARS survey). Moreover, they report the range of 0.01 to 1.6 $M_{\odot}$yr$^{-1}$ when calculating the outflow mass rate in kiloparsec scales, and values up to thirty times higher when considering a nuclear outflow radius of 27 pc. These higher values are compatible with the ones of our compact sources (see Table~\ref{tab:of-prop}) and also with the ones of the control sample of extended RG. Regarding the frequency of outflowing gas in our sample, we found it very similar to the frequency found by \citep{Woo16} in radio-weak sources (45\%). In their work, they study a large sample of 39\,000 type~2 AGN.  This is an interesting result in terms of the role of the AGN radiation in driving ionized outflows, which seems to be significant in both radio-weak and radio-loud galaxies.

\subsection{Compact sources in the life cycle of radio galaxies}

The outflows in RG are an important source of information about the early life cycle in AGN, that is, the fraction of time that the AGN is accreting matter. More specifically, the CSS and GPS galaxies, that have spectral indexes in the frequency 144~MHz $\alpha>0$, would represent the newborn jets and also a grown phase jet. In such phases, the radio jets are going out of the inner regions of the host galaxies, tracing path through and interacting strongly with the ISM. As the jet grows, this interaction between both the jet and ISM decreases.

When looking at the kinetic power of the compact sources as a function of the radio luminosity (Figure~\ref{fig:kineticpowerxradio}), we see a possible anti-correlation between these parameters. This result supports the interpretation of young jets clearing a channel through the gas, so in later stages, the evolved jet has less gas to interact with. Exploring this point, we calculate the Spearman rank-order correlation coefficient and the p-value of the correlation. The Spearman coefficient measures the direction and strength of an association between two parameters and varies from -1 to 1, where 0 represents no correlation. We estimate the value of $\sim$-0.538, where the negative indicates an anti-correlation. The regression in kinetic power when increasing the bolometric luminosity would imply that the outflows are more energetic for lower luminosities. When we associate the lower luminosity with a more compact jet, we conclude that more compact jets drive more powerful outflows. This is in agreement with the previous discussion about the coupling between compact jets and their environment \citep{Mukherjee18b}. The p-value of the correlation is 0.058, which means the kinetic power and the radio luminosity are not significantly correlated. As the p-value is accurate for samples with more than 500 points, we have just a rough estimation. More sources are necessary to confirm this result.

\begin{table*}
    \centering
    \caption{Outflow gas properties. The columns show the source name, the bolometric luminosity $L_{bol}$, the radius of the outflow $R_{out}$, the gas electron density $\eta_{e}$, the outflow mass $M_{of}$, the outflow rate $\dot{M}_{of}$, the kinetic energy $E_{kin}$ and the kinetic power $\dot{E}_{kin}$.}
    \label{tab:of-prop}
    \begin{tabular}{llllllllr}
    \hline
    \textbf{Object}         & log \textbf{L$_{bol}$} &$R_{out}$ &$\eta_{e}$&log \textbf{$M_{of}$} & \textbf{$\dot{M}_{of}$}   & log \textbf{$E_{kin}$} & log \textbf{$\dot{E}_{kin}$}   \\
            & (erg s$^{-1}$) & (kpc) & (cm$^{-3}$) & (M$_{\odot}$) & (M$_{\odot}$ yr$^{-1}$) & (erg) & (erg s$^{-1}$) \\
    \hline
    4C+00.03 & 43.91 & 2.93 & 164 & 5.63 & 0.16 & 54.74 & 39.79   \\
    4C+07.22 & 44.23 & 3.18 & 221 & 5.30 & 0.16 & 55.11 & 40.51   \\
    B3-0833+44 & 44.13 & 1.68 & 352 & 5.41 & 0.36 & 54.92 & 40.81   \\
    WISEAJ085323.42 & 44.64 & 3.27 & 233 & 6.55 & 4.05 & 56.37 & 42.18   \\
    B3-0902+46 & 43.60 & 2.48 & 300 & 5.63 & 0.58 & 55.53 & 41.35   \\
    J0945+1737 & 45.68 & 3.56 & 300 & 7.30 & 9.63 & 56.64 & 41.97   \\
    FBQSJ09452 & 44.65 & 5.27 & 300 & 6.06 & 0.24 & 55.24 & 39.97   \\
    J0958+1439 & 45.76 & 3.11 & 593 & 6.49 & 2.20 & 56.28 & 41.54   \\
    J1000+1242 & 45.39 & 4.03 & 300 & 7.40 & 10.64 & 56.83 & 42.01   \\
    J1010+1413 & 45.93 & 5.10 & 300 & 7.53 & 24.65 & 57.60 & 43.05   \\
    J1010+0612 & 44.74 & 2.83 & 391 & 6.33 & 2.00 & 56.25 & 41.66   \\
    J1100+0846 & 45.47 & 2.89 & 300 & 6.83 & 3.53 & 56.23 & 41.42   \\
    B3-1154+435 & 46.13 & 5.70 & 300 & 7.06 & 4.50 & 56.77 & 41.88   \\
    B3-1241+411 & 45.19 & 6.05 & 300 & 7.25 & 7.02 & 56.99 & 42.12   \\
    J1338+1503 & 45.50 & 4.84 & 300 & 6.99 & 2.75 & 56.05 & 41.22   \\
    4C+12.50 & 45.00 & 3.41 & 433 & 6.87 & 15.03 & 57.40 & 43.37   \\
    4C+62.22 & 44.58 & 8.66 & 300 & 5.83 & 12.60 & 55.48 & 42.23   \\
    2MASXj1511 & 44.41 & 2.46 & 300 & 6.26 & 2.15 & 56.14 & 41.78   \\
    {[}HB89{]}2247 & 45.24 & 2.32 & 300 & 6.81 & 1.45 & 56.00 & 40.84    \\
    \hline
    \end{tabular}
\end{table*}

\subsection{Kinetic power and feedback efficiency}

When the ratio between kinetic power and bolometric luminosity is above 0.5\%, the coupling between the feedback and the ISM is efficient. Depending on the velocity used in Equation \ref{eq:kineticpower}, the number of sources above the value of $\dot{E}_{kin}$/$L_{bol}$= 0.01 change, but the behavior of the parameters does not change. With the velocity $v_{50}$, we have one source with $\sim$0.2\% of feedback efficiency, with a mean value of 0.09\% and a lower limit of 0.002\% for the sample. Using the velocity as $v_{max}$, it changes less than 1 dex in each plot. Also, the number of sources with high feedback efficiency increases to four, and the mean value is 0.4\%. This is a higher value in comparison to what was obtained for CSS of 0.13\% in a stacking procedure by \cite{Kukreti23} (hereafter K23), and also higher than that found in extended RG of 0.2\% by \cite{Speranza21}.

We note that K23 found lower values of outflow rates. However, they used a stacking procedure to detect outflows in lower S/N objects. They used a mean value of the electron density for all their sources, as well as the mean value of the radius of the region where the outflow is taking place. This is valuable for understanding the behavior of the CSS as a group. In our work, we are studying individual sources, so we estimate individual values for the density and radius of outflow gas. We saw that these values have a large range because of the different redshift of each source. Computing these parameters for each object allows us to study compact sources individually and the variation in the different characteristics. For example, the detection of three different types of  \oiii{} profiles reflects the different nature of the outflowing gas. In some cases, it could be accompanied by the formation and expansion of gas cocoons as the jet propagates. This scenario would explain the observation of split line profiles found in 10\% of the sample. Moreover, Figure~\ref{fig:deltav-fwhm} tells us that in nearly half the sample where a broad component in \oiii{} is detected, the line centroid is redshifted to the systemic velocity. 

\cite{Kukreti23} found broader \oiii{} profiles in peaked sources than in non-peaked sources. Using the relation between the radio spectral peak and the age of the radio jet, their results point out that younger sources have more disturbed ionized gas. Similarly, as we see from our results for individual compact galaxies, more compact jets, believed to be in the early stages of RG \citep{ODea21}, tend to drive more powerful ionized outflows. The physical parameters estimated here corroborate with the conclusions of K23 and other earlier findings \citep{Holt08,Mullaney13,Zovaro19} of the strong impact of the radio jets in the ISM in their young stages.

\subsubsection{Relation to stellar parameters}

We also investigated the possible relation between the outflow properties with the stellar parameters, such as the stellar mass and age of the stellar population. We found no relation in such a comparison. This indicates that there is no positive feedback in our sample, but a study with a greater number of young sources is necessary to conclude if it is a general result for CSS and GPS.

\subsubsection{Relation to BH mass and LS}

%\section{Black hole mass versus linear size}
%\label{sec:ap:bhmass-ls}

\begin{figure}
\begin{center}

\includegraphics[width=\columnwidth]{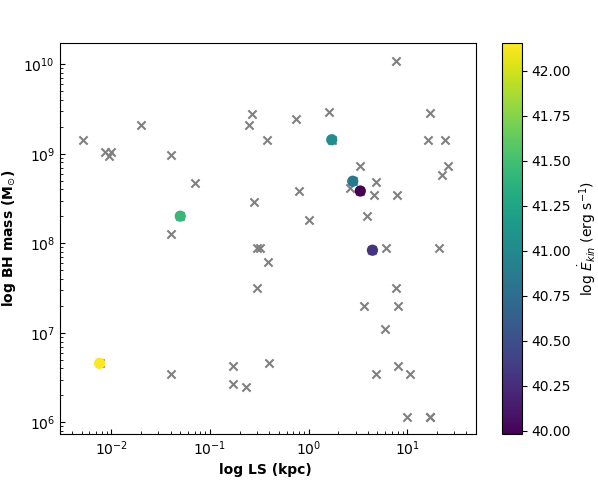} 
\caption{Black hole mass versus linear size (LS) of compact galaxies in our sample. The color scale represents sources with measured kinetic power.   }
\label{fig:bhmass-linearsize}
\end{center}
\end{figure}

To investigate the relationship between black hole mass and the linear size of radio emission in compact galaxies, we plot these parameters and also consider the kinetic power to determine if any trends are present, as shown in Figure~\ref{fig:bhmass-linearsize}. If more compact jets interact more extensively with the interstellar medium (ISM), the outflows could potentially be more powerful. Our analysis indicates that there is no specific trend between black hole mass and linear size for our sample. Furthermore, additional data on outflow kinetic power is needed to confirm any potential anticorrelation.

We see from the BH mass calculation that 40\% of our sample have BH masses lower than those of extended RG ($<10^8M_{\odot}$). When plotting the LS versus BH masses of our sample, we see no significant trend between these parameters (see Figure~\ref{fig:bhmass-linearsize}). However, adding the kinetic power of the outflow in this plot, we see sources driving outflows with BH masses lower than 10$^{8}$~M$_{\odot}$. As discussed in Section~\ref{sec:bhmass-estimation}, sources with lower masses are not expected to evolve into extended RG, however, they still produce powerful outflows comparable to those of extended objects.

\section{Conclusions}
\label{sec:conclusions}

We have performed a study of optical SDSS-DR12 spectra of 82 compact radio galaxies classified as CSS, GPS, and MPS. We aim to characterize the ionized gas of these objects and analyze the properties of outflowing gas defined by the detection of broad \oiii{}~$\lambda5007$ emission in the individual spectra. We fitted Gaussian profiles in the strongest emission lines to obtain the fluxes and then the physical parameters of the ionized gas. From our analysis, we conclude the following. 

\begin{itemize}
    \item Almost half (43\%) of our sample present broad emission in the \oiii{}~$\lambda5007$ line, which we interpret as due to outflowing gas in the nuclear region of compact galaxies. This frequency is smaller than the ones reported in the literature for extended RG and radio-weak AGN. 

    \item When comparing our results to the ones in the literature, we see that compact jets drive outflows as massive as the nuclear ones of extended RGs, with similar outflow rates and kinetic energy. This reinforces the prediction of stronger jet-ISM coupling in the inner stages of jets. Also, the median value of the kinetic efficiency of compact galaxies is higher than the nuclear values reported for extended RGs.  
    
    \item No evidence of positive feedback was found in our data when we studied the possible correlation between outflow and stellar mass and age.

    \item CLs are observed in 18 out of our sample of 82 galaxies, representing 22\% of the compact galaxies. This occurrence is lower compared to the frequency observed in radio-weak AGN, where coronal lines are present in 45\% to 64\% of the galaxies. This reflects the strong variation in gas excitation in compact galaxies when compared to classical AGN. For the compact galaxies where the \rm{[Fe \sc{vii}]}/\oiii{} ratio between the narrow components was measured, we found a mean value of 0.04. That value is very similar to the one displayed by radio-weak AGN, which ranges between 0.01 and 0.08. It means that the ratio of these two ions is probably not related to the jet or is not jet-excited but produced by radiation from the AGN. However, the broad components are likely jet-excited.

    \item The kinetic power decreases with the radio luminosity in the range up to 10$^{25}$~W\,Hz$^{-1}$, showing an anti-correlation that needs to be investigated with more compact objects. The values of linear size for our compact galaxies are scarce, so a relationship between outflow parameters and the linear size of the jet was not evident.

\end{itemize}

To summarise, compact radio galaxies can drive powerful outflows of gas. The rate and power of the outflows are comparable to the nuclear values reported for extended RG, although it seems to be decreasing with linear size. In this sense, our work provides observational insights about the impact of compact jets in ISM and reinforces the need to study outflows individually and at larger spectral and angular resolutions.

\section{Acknowledgments}

We would like to thank Giovanna Speranza for the data provided.  This study was financed in part by the Coordena\c{c}\~{a}o de Aperfei\c{c}oamento de Pessoal de N\'{i}vel Superior - Brazil (CAPES) - Finance Code 001 (B.L.M.M.) and Conselho Nacional de Desenvolvimento Científico e Tecnológico (CNPq). ARA acknowledges Conselho Nacional de Desenvolvimento Científico e Tecnológico (CNPq) for partial support to this work under grant 313739/2023-4. This work has made use of the computing facilities available at the Instituto Nacional de Pesquisas Espaciais (INPE). SP is supported by the international Gemini Observatory, a program of NSF NOIRLab, which is managed by the Association of Universities for Research in Astronomy (AURA) under a cooperative agreement with the U.S. National Science Foundation, on behalf of the Gemini partnership of Argentina, Brazil, Canada, Chile, the Republic of Korea, and the United States of America. This research has made use of the VizieR catalog access tool, CDS, Strasbourg, France (DOI: \url{10.26093/cds/vizier}). The original description of the VizieR service was published in A\&AS 143, 23. This research has made use of NASA’s Astrophysics Data System.

Funding for SDSS-III has been provided by the Alfred P. Sloan Foundation, the Participating Institutions, the National Science Foundation, and the U.S. Department of Energy Office of Science. The SDSS-III web site is \url{http://www.sdss3.org/}.

SDSS-III is managed by the Astrophysical Research Consortium for the Participating Institutions of the SDSS-III Collaboration including the University of Arizona, the Brazilian Participation Group, Brookhaven National Laboratory, Carnegie Mellon University, University of Florida, the French Participation Group, the German Participation Group, Harvard University, the Instituto de Astrofisica de Canarias, the Michigan State/Notre Dame/JINA Participation Group, Johns Hopkins University, Lawrence Berkeley National Laboratory, Max Planck Institute for Astrophysics, Max Planck Institute for Extraterrestrial Physics, New Mexico State University, New York University, Ohio State University, Pennsylvania State University, University of Portsmouth, Princeton University, the Spanish Participation Group, University of Tokyo, University of Utah, Vanderbilt University, University of Virginia, University of Washington, and Yale University.

\vspace{5mm}
\facilities{SDSS}

\software{numpy \citep{harris2020numpy},
          scipy \citep{2020SciPy-NMeth},
          matplotlib \citep{Hunter2007matplotlib},
          PySpecKit \citep{Ginsburg11}
          }

\bibliography{sample631}{}
\bibliographystyle{aasjournal}

\appendix

\section{Properties and emission line fluxes of the compact sources.}
\label{sec:appendixA}

We show in Table~\ref{tab:general-properties} the general properties, such as coordinates, radio luminosity (L$_r$), linear size (LS), distance and scale of the compact sources in our sample. The obtained electron density from the sulfur emission lines is also shown. In Table~\ref{tab:bhmass-sigma} we show the stellar velocity dispersion $\sigma_{*}$ and the black hole mass obtained in our analysis. 

Table~\ref{tab:flux-emission-lines} and Table~\ref{tab:flux-emission-lines2} list the fluxes and FWHM of the lines that are relevant to this work, that is, H$\beta$, \oiii{}~$\lambda$4959,  \oiii{}~$\lambda$5007, H$\alpha$, \rm{[N \sc{ii}]}~$\lambda$6584, \rm{[S \sc{ii}]}~$\lambda$6716, and \rm{[S \sc{ii}]}~$\lambda$6732. The fluxes are given in units of  10$^{-15}$ erg \, s$^{-1}$cm$^{-2}$\AA$^{-1}$ and the FWHM in \AA. When more than one component was fit to a given line we provide the parameters for each component of the line. 

%======================================
\begin{longtable}{lcccccccr}
\caption{Properties of the compact sources in our sample.} 
\label{tab:general-properties}\\
\hline
Source & RA & Dec & z & Log L$_r$ &  LS$^{*}$ &  Scale$^{**}$ & $\eta_{e}$\\
      &     &     &   & (WHz$^{-1}$) &   (kpc) &  (kpc\,arcsec$^{-1}$) & (cm$^{-3}$)\\
\hline\hline
\endfirsthead 
\caption[]{Continuing} \\
\hline
Source & RA & Dec & z & Log L$_r$ &  LS$^{*}$   & Scale$^{**}$ & $\eta_{e}$\\
      &     &     &   & (WHz$^{-1}$) &   (kpc) &  (kpc\,arcsec$^{-1}$) & (cm$^{-3}$)\\
\hline\hline
\endhead
\endlastfoot
\hline
\endfoot
4C+00.02	&	00:22:25.42	&	00:14:56.06	&	0.306	&	26.64	&		0.27	&	6.57	&	164 \\
4C+00.03	&	00:28:33.42	&	00:55:11.02	&	0.104	&	24.71	&		3.35	&	1.96	&	237 \\
B2-0026+34	&	00:29:14.20	&	34:56:32.00	&	0.517	&	26.81	&		--	&	11.09	&	300\\
3C049	&	01:41:09.10	&	13:53:28.00	&	0.620	&	27.37	&		--	&	13.32	&	300 \\
B3-0754+401	&	07:57:56.69	&	39:59:36.01	&	0.066	&	23.95	&		0.25	&	1.41	&	232 \\
4C+47.27	&	08:04:13.91	&	47:04:43.07	&	0.510	&	26.70	&		6.18	&	10.93	&	300 \\
4C+40.20	&	08:12:53.13	&	40:19:00.09	&	0.551	&	26.6	&		7.72	&	6.61	&	300 \\
4C+07.22	&	08:13:23.76	&	07:34:05.76	&	0.112	&	25.06	&		2.81	&	2.12	&	221 \\
\end{longtable}

\begin{longtable}{lccccc}
\caption{Stellar velocity dispersion $\sigma_{*}$ and black hole mass of the compact sources from our analysis} 
\label{tab:bhmass-sigma}\\
\hline
Name          & $\sigma_{*}$ & log M$_{BH}$  & Name           & $\sigma_{*}$ & log M$_{BH}$      \\
                 & ($km\,s^{-1})$ &  ($M_{\odot}$)  &                   & $(km\,s^{-1})$ &  ($M_{\odot}$)      \\
\hline\hline
\endfirsthead 
\caption[]{Continuing} \\
\hline
Name            & $\sigma_{star}$ & log M$_{BH}$  &  Name            & $\sigma_{*}$ & log M$_{BH}$      \\
                 & ($km\,s^{-1})$ &  ($M_{\odot}$)  &                   & $(km\,s^{-1})$ &  ($M_{\odot}$)      \\
\hline\hline
\endhead
\endlastfoot
\hline
\endfoot
J0945+1737	&	163.69$\pm$24.55	&	8.11$\pm$1.22	&	4C+14.41	&	76.4$\pm$11.46	&	6.66$\pm$1.00	\\
SDSSJ08531	&	93.5$\pm$14.03	&	7.04$\pm$1.06	&	B3-1308+451	&	283.65$\pm$42.55	&	9.15$\pm$1.37	\\
FBQSJ07411	&	150$\pm$22.50	&	7.94$\pm$1.19	&	2MASXiJ140	&	219.85$\pm$32.98	&	8.67$\pm$1.30	\\
4C+47.27	&	150$\pm$22.50	&	7.94$\pm$1.19	&	SDSS091734	&	71.68$\pm$10.75	&	6.54$\pm$0.98	\\
J1000+1242	&	222.55$\pm$33.38	&	8.69$\pm$1.30	&	2MASXJ0834	&	264.05$\pm$39.61	&	9.02$\pm$1.35	\\
SDSSJ14210	&	150$\pm$22.50	&	7.94$\pm$1.19	&	B3-1128+455	&	222.12$\pm$33.32	&	8.69$\pm$1.30	\\
J0958+1439	&	85.13$\pm$12.77	&	6.87$\pm$1.03	&	B3-1241+411	&	181.29$\pm$27.19	&	8.30$\pm$1.25	\\
2MASSJ0934	&	334.77$\pm$50.22	&	9.47$\pm$1.42	&	B3-1445+410	&	243.62$\pm$36.54	&	8.87$\pm$1.33	\\
\end{longtable}

\begin{longrotatetable}
\tiny
\centering
\begin{longtable}{lcccccccccccc}
\caption{Fluxes (in units of 10$^{-15}$ erg \, s$^{-1}$cm$^{-2}$\AA$^{-1}$) and FWHM (\AA) of the emission lines employed in our analysis.} 
\label{tab:flux-emission-lines}\\
\hline
Name    & H$\beta$ & FWHM & H$\beta$b & FWHM & [O \sc{iii}] $\lambda$4959 & FWHM & [O \sc{iii}]$\lambda$4959b & FWHM & [O \sc{iii}]$\lambda$5007 & FWHM & [O \sc{iii}]$\lambda$5007b  & FWHM \\

\hline\hline
\endfirsthead 
\caption[]{Continuing} \\
\hline
Name    & H$\beta$ & FWHM & H$\beta$b & FWHM & [O \sc{iii}] $\lambda$4959 & FWHM & [O \sc{iii}]$\lambda$4959b & FWHM & [O \sc{iii}]$\lambda$5007 & FWHM & [O \sc{iii}]$\lambda$5007b  & FWHM \\

\hline\hline
\endhead
\endlastfoot
\hline
\endfoot
4C+00.02           & 0.08$\pm$0.01 & 12.56  & 0.34$\pm$0.05  & --          & 0.15$\pm$0.02  & 8.88     & --             & --          & 0.46$\pm$0.07  & 8.88     & --             & --          \\
4C+00.03           & 2.19$\pm$0.33 & 6.44   & 5.87$\pm$0.88  & --          & 2.05$\pm$0.31  & 3.66     & 2.83$\pm$0.42  & 10.73         & 6.16$\pm$0.92  & 3.66     & 8.51$\pm$1.28  & 10.73         \\
B2-0026+34         & 0$\pm$0       & --          & --             & --          & 0.08$\pm$0.01  & 9.04     & --             & --          & 0.24$\pm$0.04  & 9.04     & --             & --          \\
3C049              & 0$\pm$0       & --          & --             & --          & --             & --          & --             & --          & --             & --          & --             & --          \\
WISEAJ0235    & 0.18$\pm$0.03 & 8.72   & 0.81$\pm$0.12  & --          & 0.11$\pm$0.02  & 8.57     & --             & --          & 0.34$\pm$0.05  & 8.57     & --             & --          \\
FBQSJ07411         & 140$\pm$21    & 57.25  & --             & 270.1       & 8.06$\pm$1.21  & 5.28     & 12.5$\pm$1.88  & 19.3          & 24.2$\pm$3.63  & 5.28     & 37.6$\pm$5.64  & 19.3          \\
B3-0754+401        & 2.29$\pm$0.34 & 6.98   & 7.61$\pm$1.14  & --          & 3.25$\pm$0.49  & 4.44     & 2.87$\pm$0.43  & 8.57          & 9.76$\pm$1.46  & 4.44     & 8.61$\pm$1.29  & 8.57          \\
4C+47.27           & 0$\pm$0       & --          & --             & --          & --             & --          & --             & --          & --             & --          & --             & --          \\
\end{longtable}
\end{longrotatetable}

\begin{longrotatetable}

\tiny
\begin{longtable}{lcccccccccccc}
\caption{Fluxes (in units of 10$^{-15}$ erg \, s$^{-1}$cm$^{-2}$\AA$^{-1}$) and FWHM (\AA) of the emission lines employed in our analysis. The fluxes are in units of $10^{-15}\,$erg\,s$^{-1}$cm$^{-2}$\AA$^{-1}$} 
\label{tab:flux-emission-lines2}\\
\hline
Name    & \rm{[N \sc{ii}]}$\lambda6548$ & FWHM & H$\alpha$ & FWHM & H$\alpha$b & FWHM  & \rm{[N \sc{ii}]}$\lambda$6584 & FWHM & \rm{[S \sc{ii}]}$\lambda6716$ & FWHM & \rm{[S \sc{ii}]}$\lambda6732$ & FWHM \\
\hline\hline
\endfirsthead 
\caption[]{Continuing} \\
\hline
Name    & \rm{[N \sc{ii}]}$\lambda6548$ & FWHM & H$\alpha$ & 
 FWHM & H$\alpha$b & FWHM  &   \rm{[N \sc{ii}]}$\lambda$6584 & FWHM & \rm{[S \sc{ii}]}$\lambda6716$ & FWHM & \rm{[S \sc{ii}]}$\lambda6732$ & FWHM \\
\hline\hline
\endhead
\endlastfoot
\hline
\endfoot

4C+00.02           & 0.11$\pm$0.02   & 11.49    & 0.34$\pm$0.05  & 12.56    & --                               & --             &          0.34$\pm$0.05 & 11.49    & 0.16$\pm$0.02                         & 11.49    & 0.13$\pm$0.02 & 11.49    \\
4C+00.03           & 3.92$\pm$0.59   & 9.71     & 5.87$\pm$0.88  & 8.7      & 7.11$\pm$1.07                    & 38.34         & 11.8$\pm$1.77 & 9.71     & 5$\pm$0.75                            & 9.71     & 4.35$\pm$0.65 & 9.71     \\
B2-0026+34         & --              & --        &          --             & --        &          --                               & --             &          --            & --        &          --                                    & --        &          --            & --        \\
3C049              & --              & --        &          --             & --        &          --                               & --             &          --            & --        &          --                                    & --        &          --            & --        \\
WISEAJ023516.81    & 0.14$\pm$0.02   & 14.05    & 0.81$\pm$0.12  & 14.08    & --                               & --             &          0.43$\pm$0.06 & 14.05    & 0.29$\pm$0.04                         & 14.05    & 0.22$\pm$0.03 & 14.05    \\
FBQSJ07411         & --              & --        &          --             & --        &          --                               & --             &          --            & --        &          --                                    & --        &          --            & --        \\
B3-0754+401        & 2.46$\pm$0.37   & 7.79     & 7.61$\pm$1.14  & 6.98     & 7.07$\pm$1.06                    & 40.43         & 7.38$\pm$1.11 & 7.79     & 3.33$\pm$0.50                         & 7.79     & 2.89$\pm$0.43 & 7.79     \\
\end{longtable}
\end{longrotatetable}

\section{Fitting of the integrated spectra}
\label{sec:apA:fitting}

In this Section we show the fitting of the emission line in the SDSS-DR12 spectra of three compact sources. The complete figure set (21 images) is available in the online journal. The blocks in left comprehend the region of \oiii{} and H$\beta$ lines, and the right panel shows the H$\alpha$ region. For some sources, the SDSS spectra do not cover the full optical range, which prevents the H$\alpha$ region from being visible in the corresponding plots.

%%%figura principal

\begin{figure*}
\begin{center}

\includegraphics[width=\textwidth]{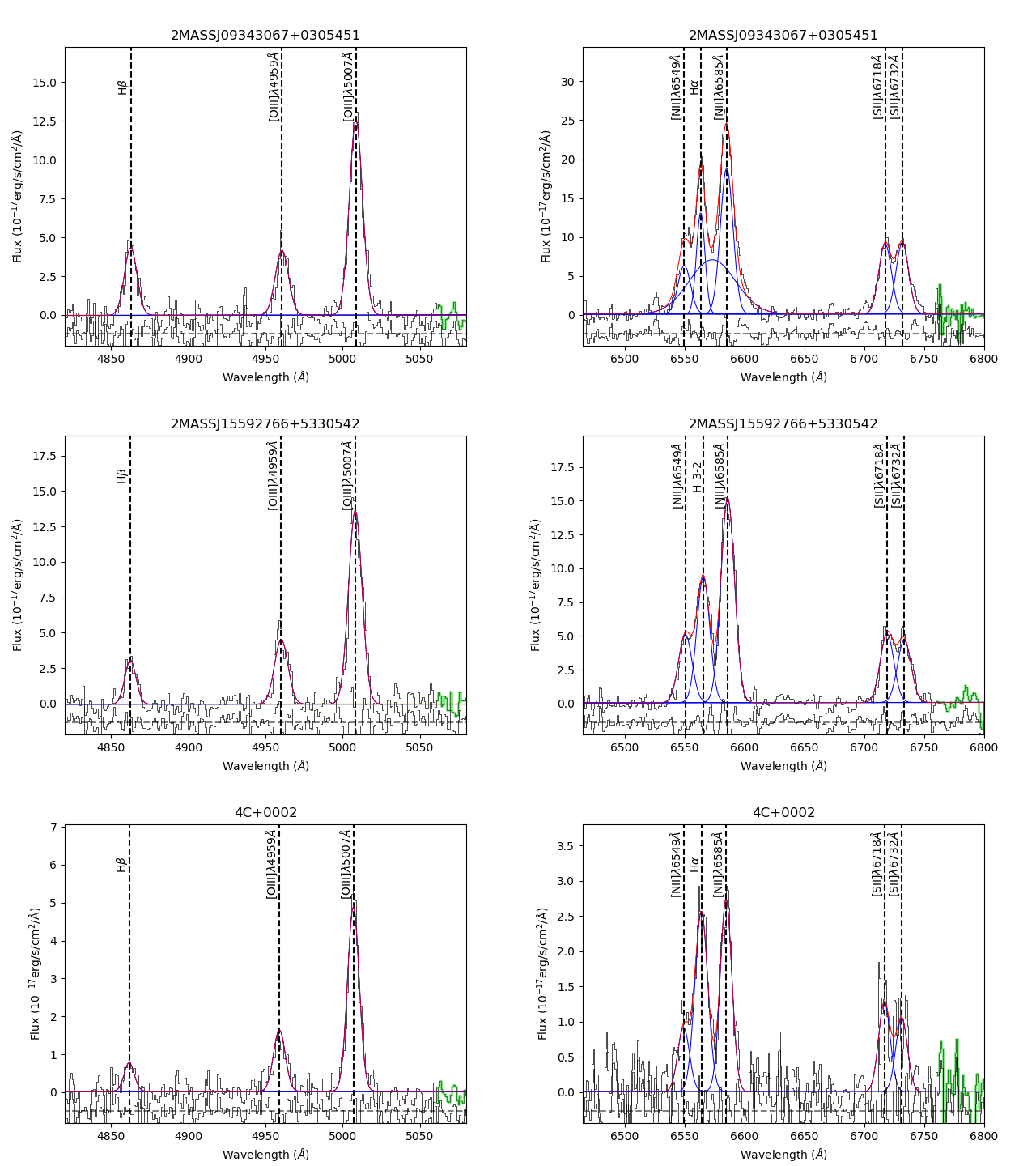}

\caption{Integrated spectra fitting of the CSS and GPS sources with the residuals.   }
\label{fig:fitting-css1}
\end{center}
\end{figure*}
%===============================================

\begin{figure*}
\begin{center}

\includegraphics[width=\textwidth]{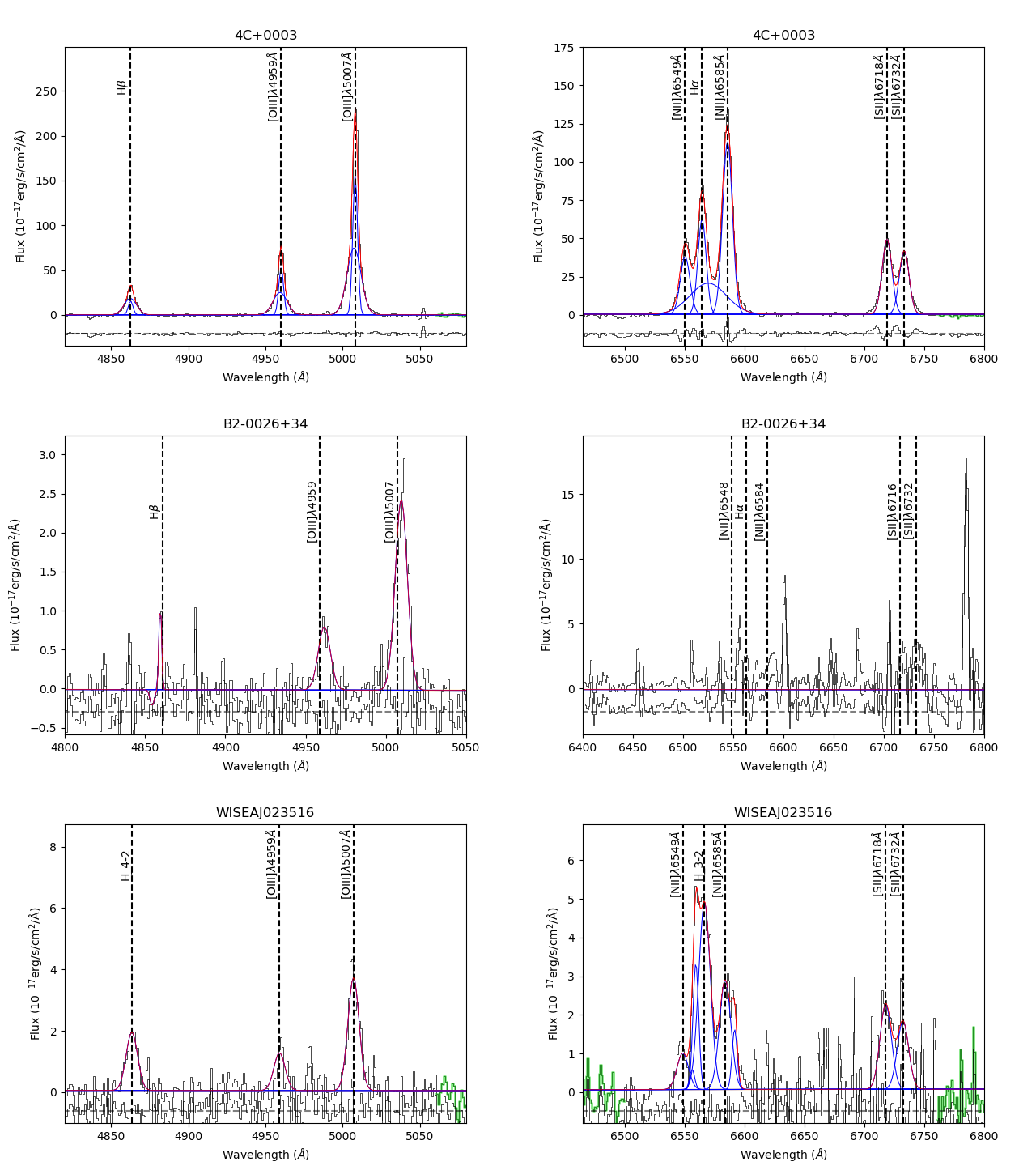} 

\caption{Integrated spectra fitting of the CSS and GPS sources with the residuals. }
\label{fig:fitting-css}
\end{center}
\end{figure*}

\begin{figure*}
\begin{center}

\includegraphics[width=\textwidth]{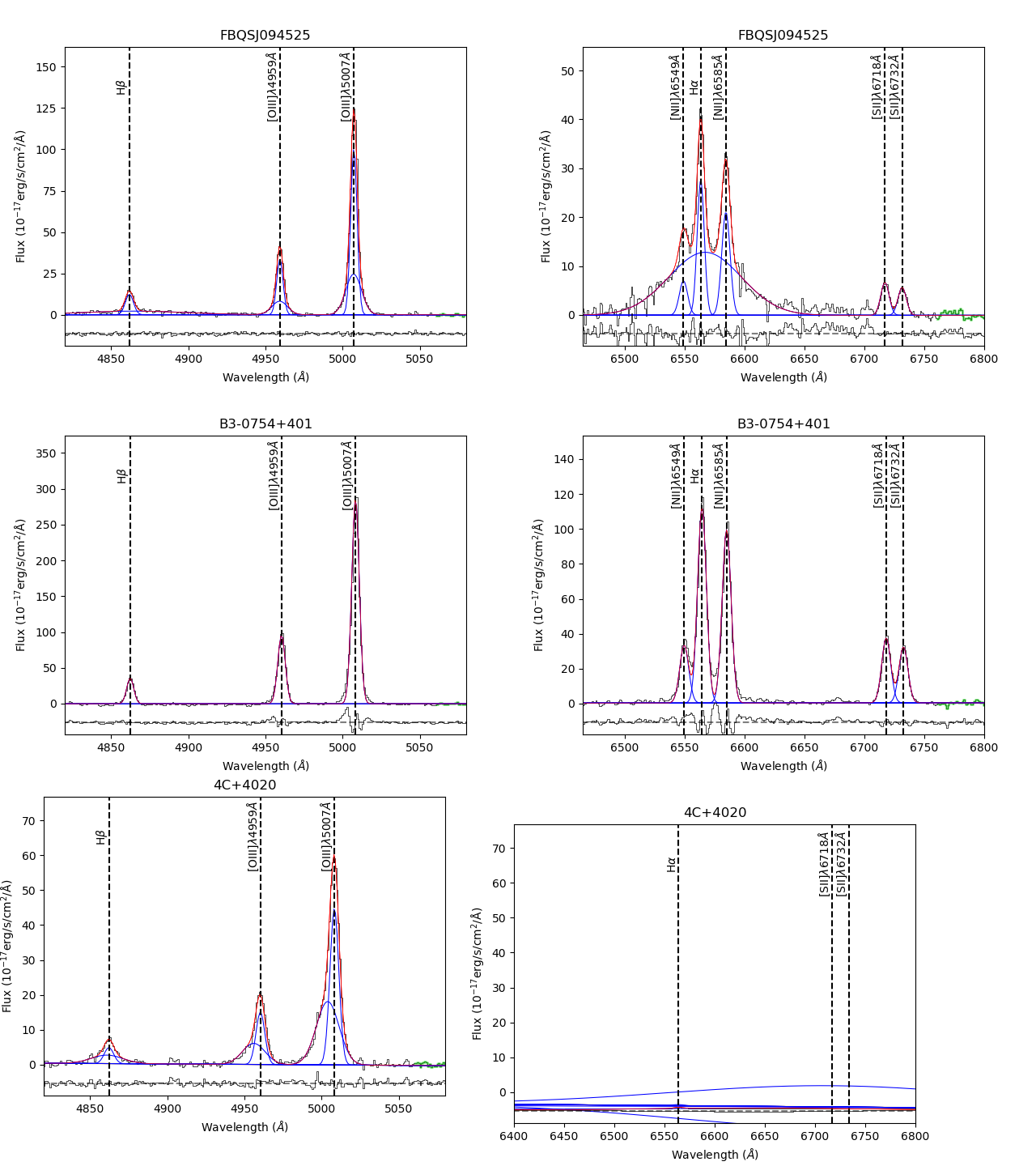}

\caption{Integrated spectra fitting of the CSS and GPS sources with the residuals. }
\label{fig:fitting-css}
\end{center}
\end{figure*}

\begin{figure*}
\begin{center}

\includegraphics[width=\textwidth]{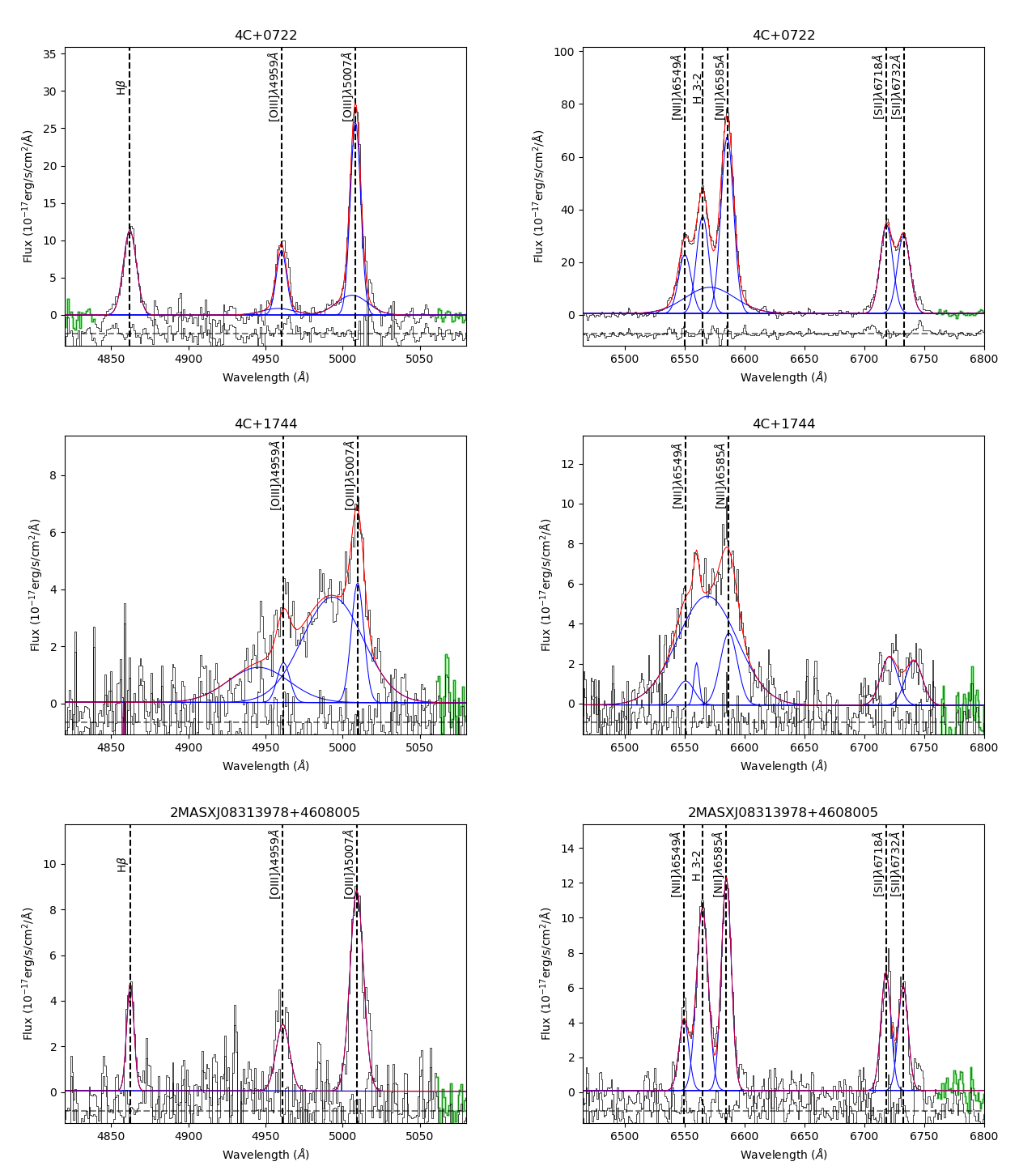}

\caption{Integrated spectra fitting of the CSS and GPS sources with the residuals.  }
\label{fig:fitting-css}
\end{center}
\end{figure*}

\begin{figure*}
\begin{center}

\includegraphics[width=\textwidth]{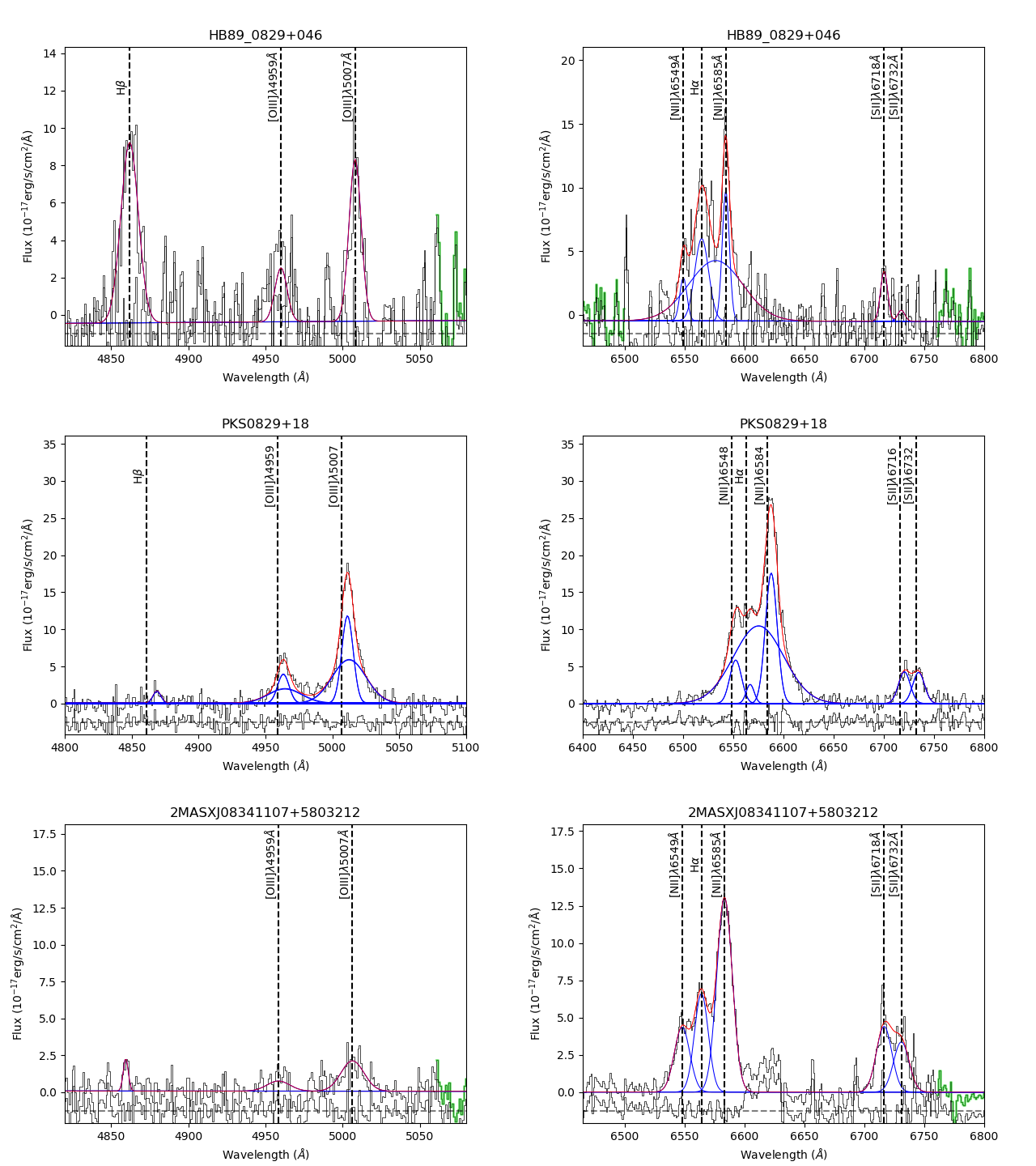}

\caption{Integrated spectra fitting of the CSS and GPS sources with the residuals.  }
\label{fig:fitting-css}
\end{center}
\end{figure*}

\begin{figure*}
\begin{center}

\includegraphics[width=\textwidth]{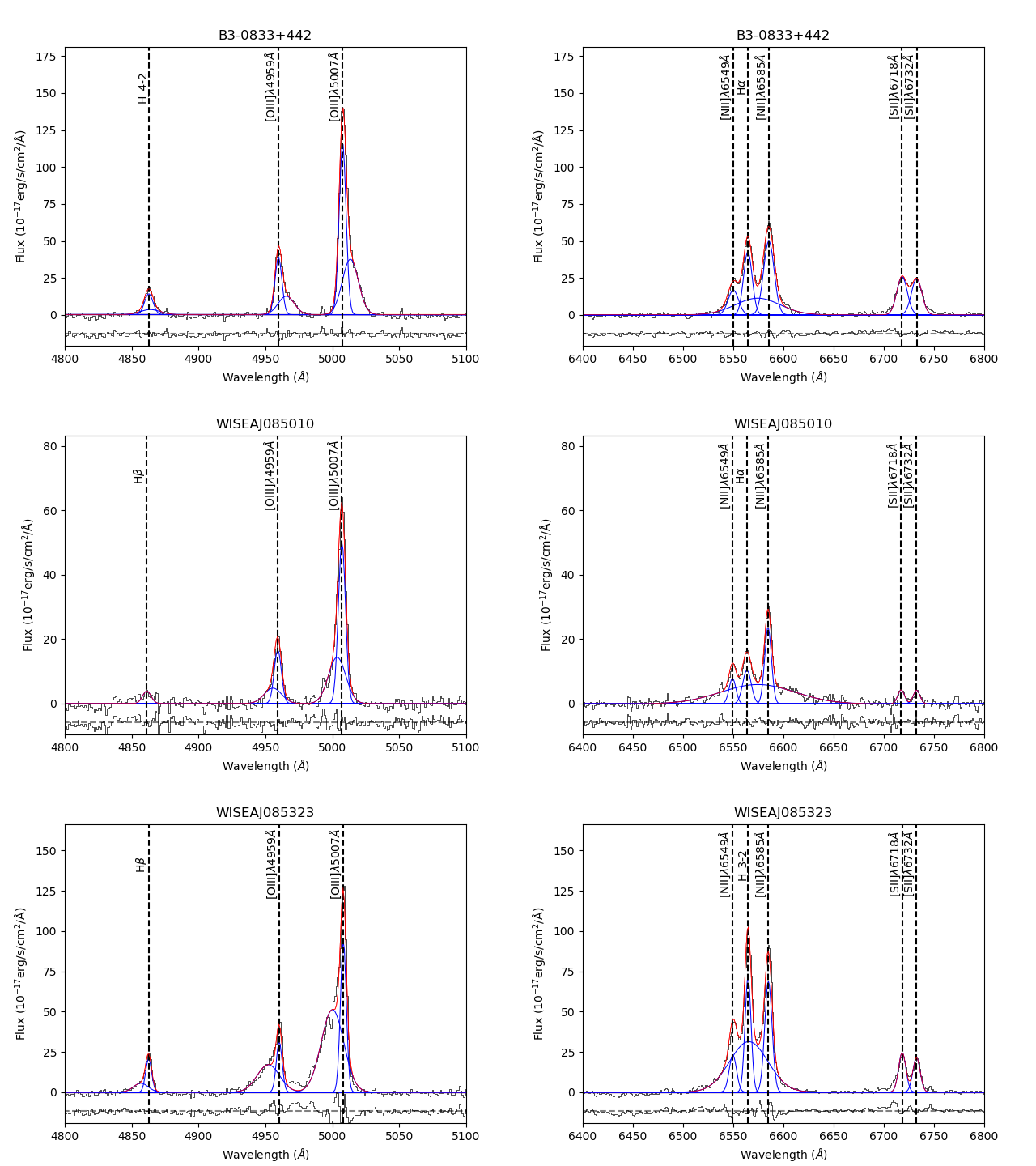}

\caption{Integrated spectra fitting of the CSS and GPS sources with the residuals.   }
\label{fig:fitting-css}
\end{center}
\end{figure*}

\begin{figure*}
\begin{center}

\includegraphics[width=\textwidth]{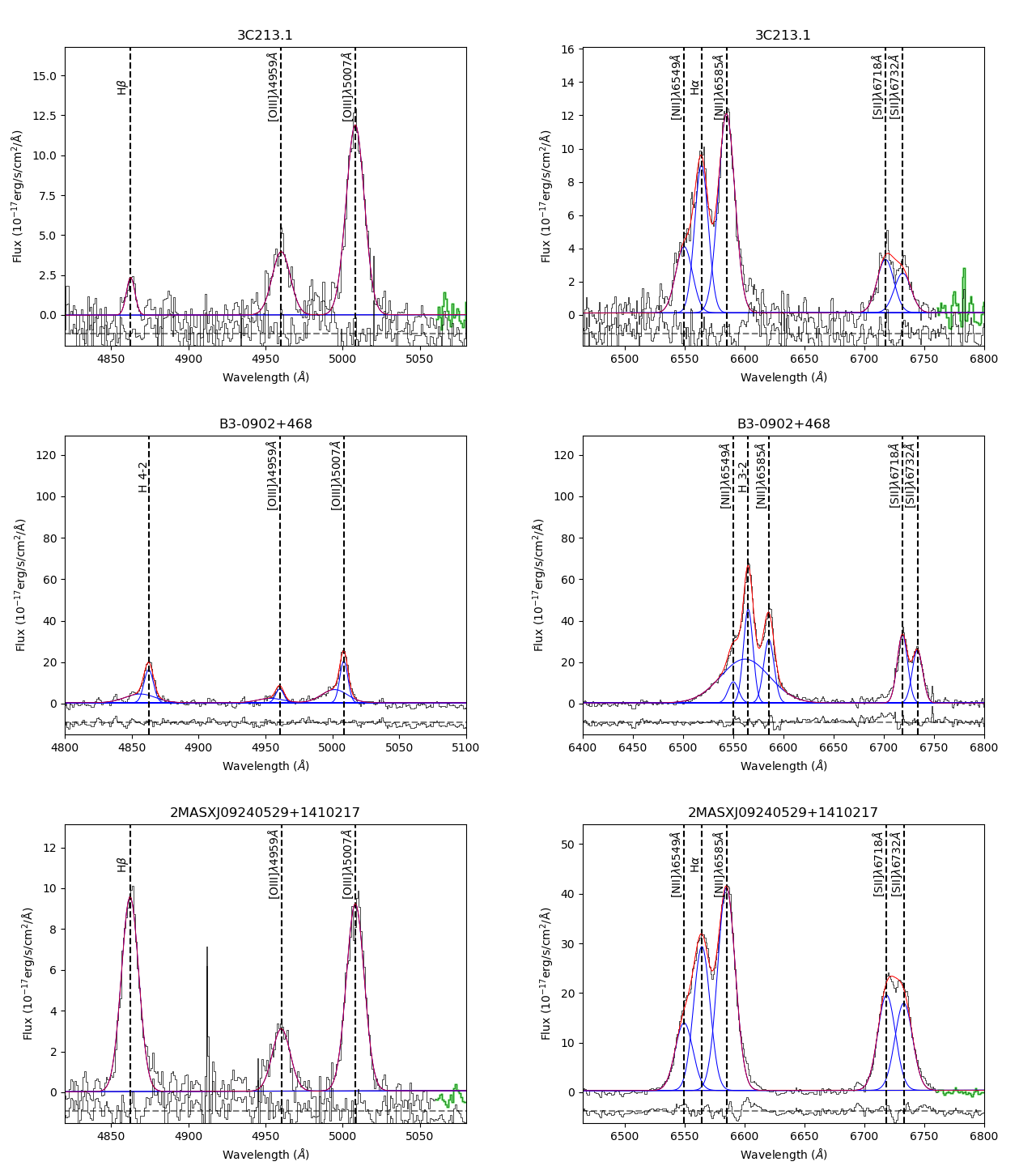}

\caption{Integrated spectra fitting of the CSS and GPS sources with the residuals.   }
\label{fig:fitting-css}
\end{center}
\end{figure*}

\begin{figure*}
\begin{center}

\includegraphics[width=\textwidth]{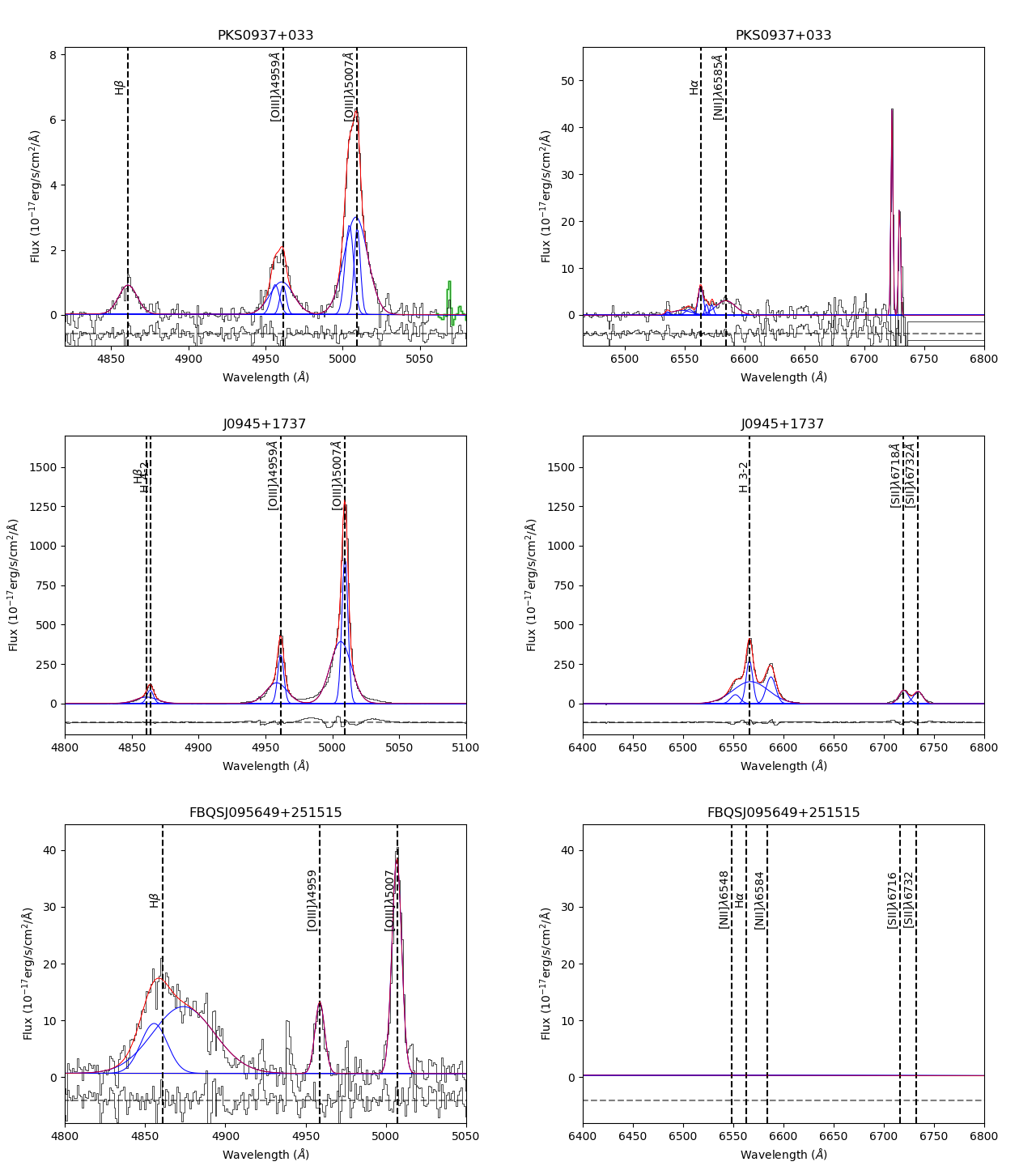} 

\caption{Integrated spectra fitting of the CSS and GPS sources with the residuals.   }
\label{fig:fitting-css}
\end{center}
\end{figure*}

\begin{figure*}
\begin{center}

\includegraphics[width=\textwidth]{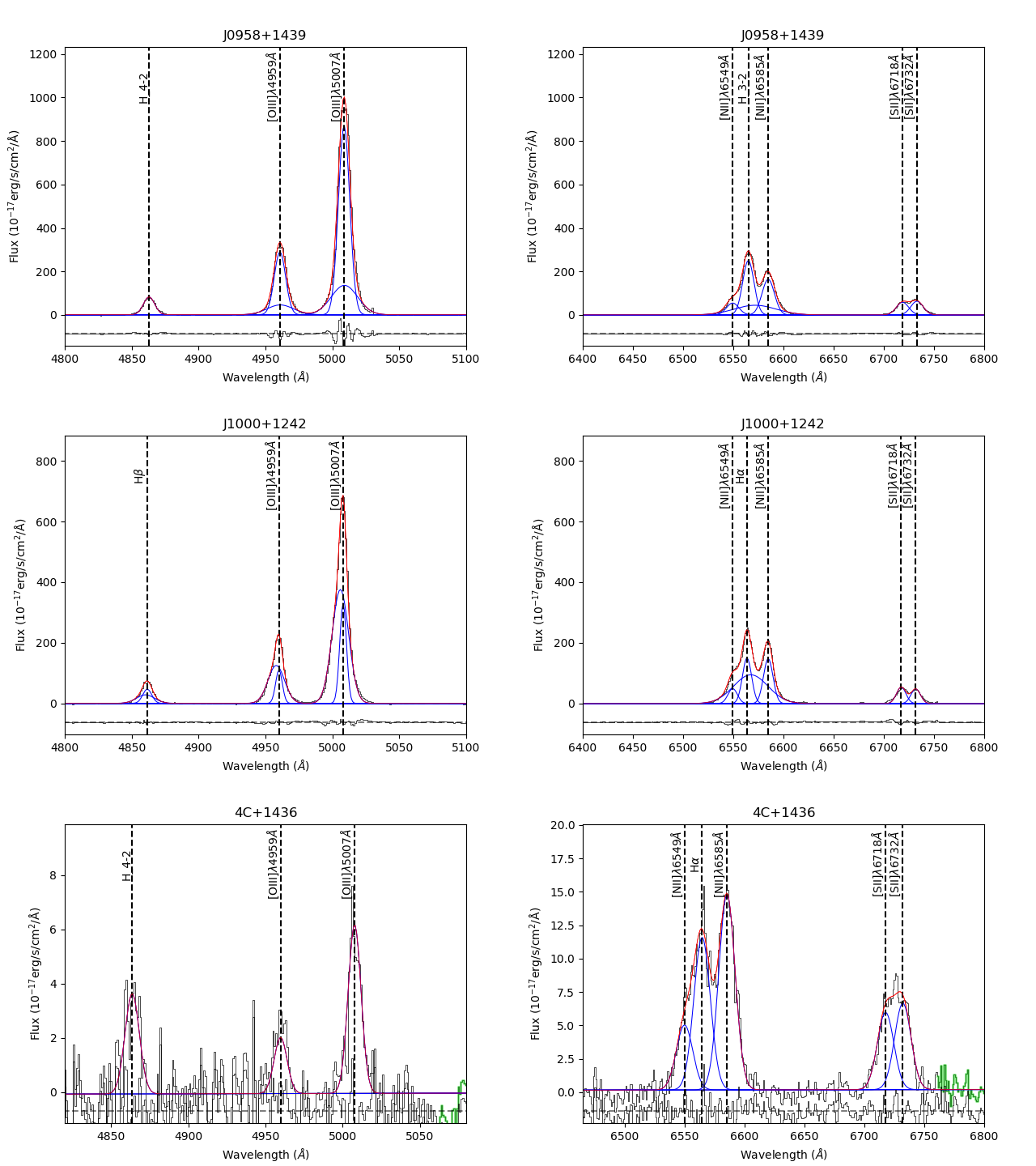}

\caption{Integrated spectra fitting of the CSS and GPS sources with the residuals.   }
\label{fig:fitting-css}
\end{center}
\end{figure*}

\begin{figure*}
\begin{center}

\includegraphics[width=\textwidth]{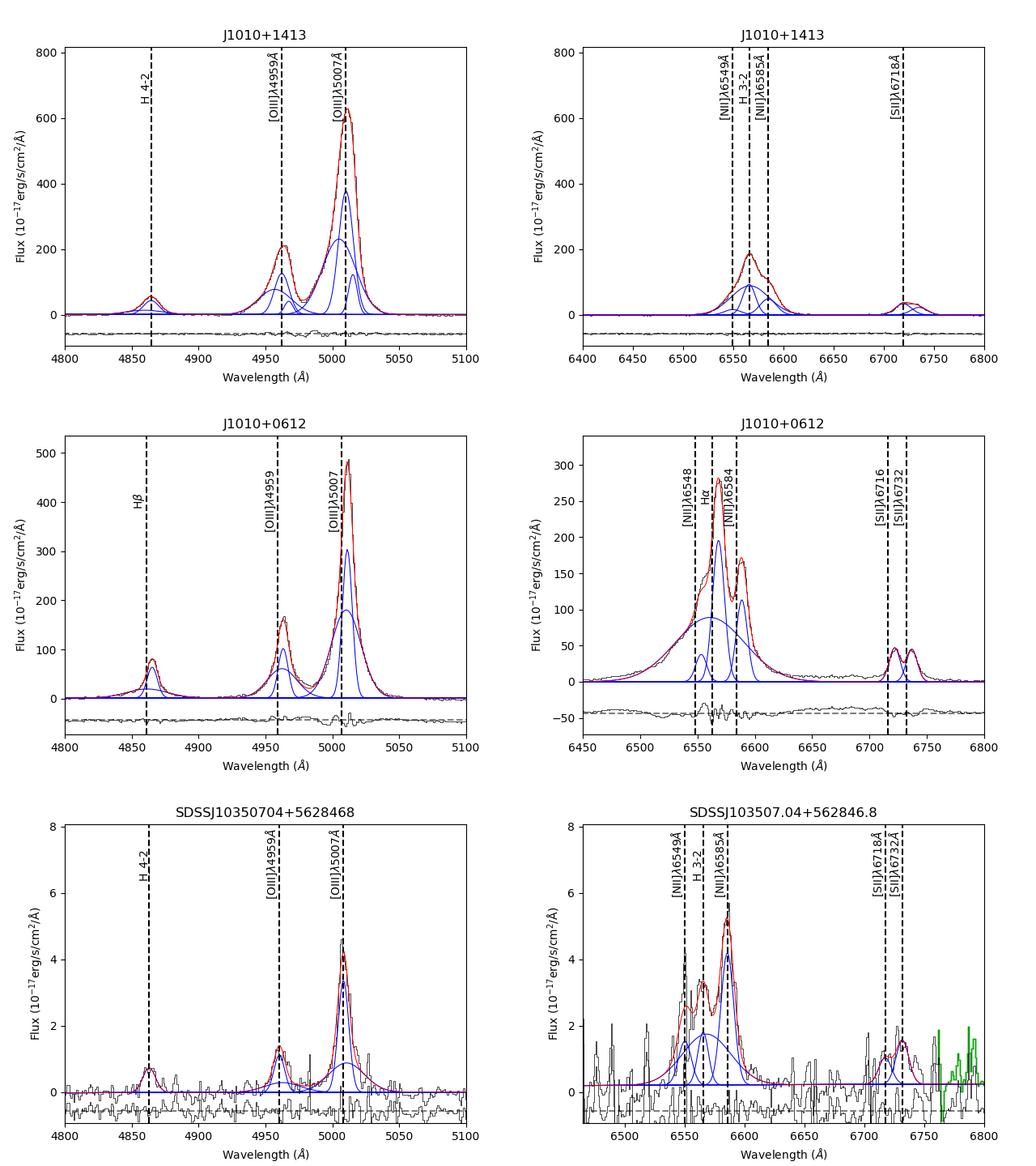}

\caption{Integrated spectra fitting of the CSS and GPS sources with the residuals.   }
\label{fig:fitting-css}
\end{center}
\end{figure*}

\begin{figure*}
\begin{center}

\includegraphics[width=\textwidth]{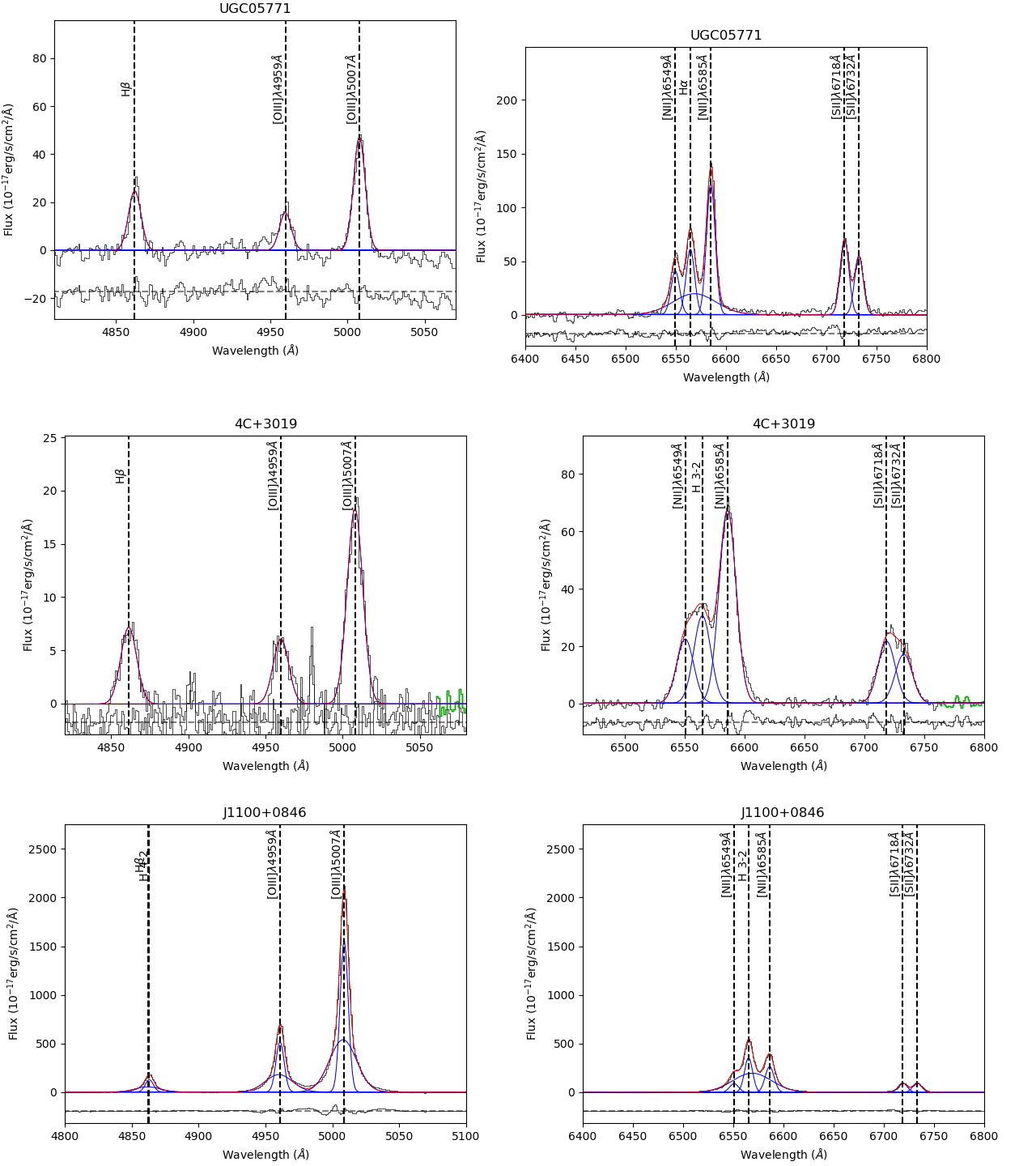}

\caption{Integrated spectra fitting of the CSS and GPS sources with the residuals.   }
\label{fig:fitting-css}
\end{center}
\end{figure*}

\begin{figure*}
\begin{center}

\includegraphics[width=\textwidth]{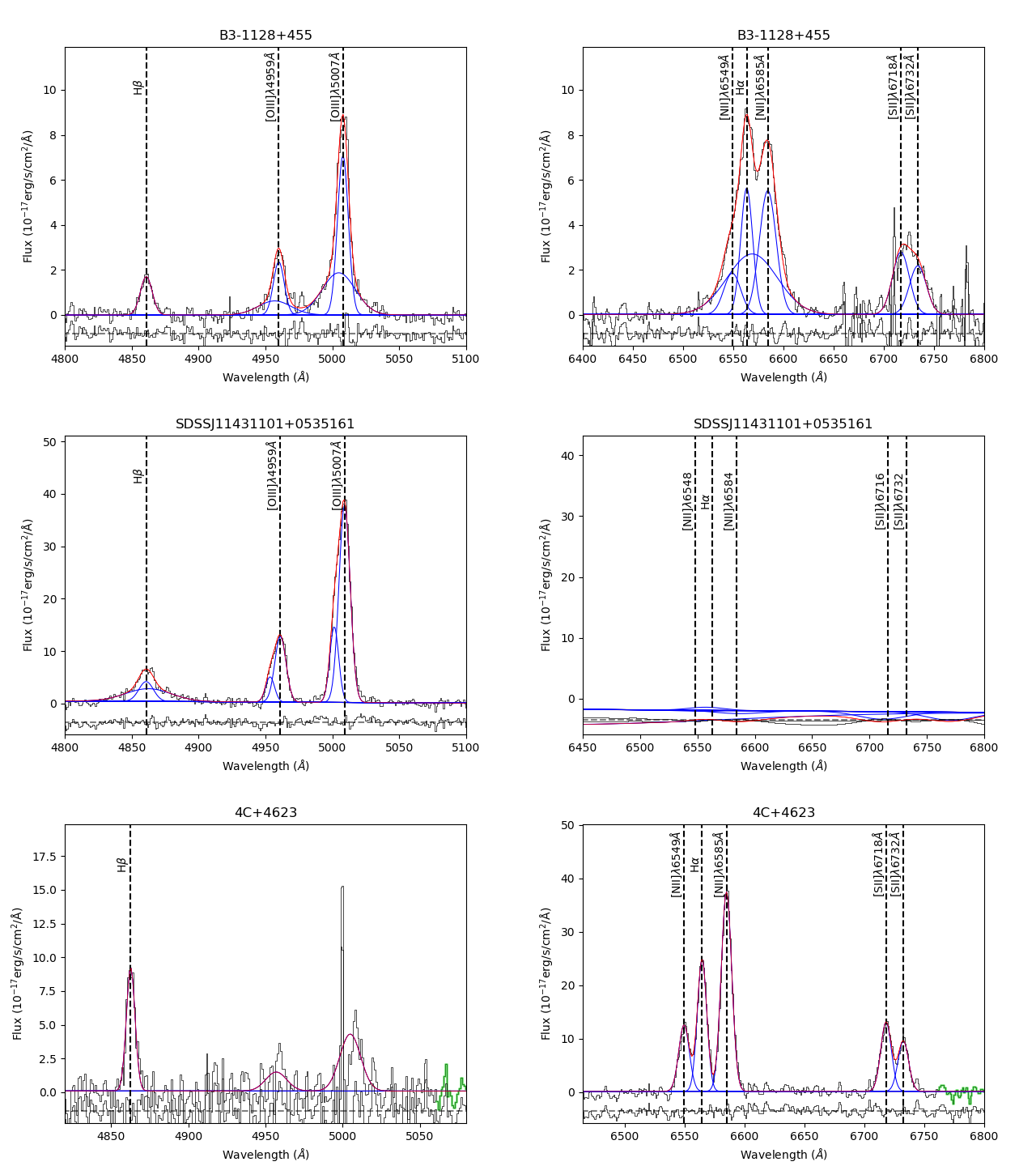}

\caption{Integrated spectra fitting of the CSS and GPS sources with the residuals.   }
\label{fig:fitting-css}
\end{center}
\end{figure*}

\begin{figure*}
\begin{center}

\includegraphics[width=\textwidth]{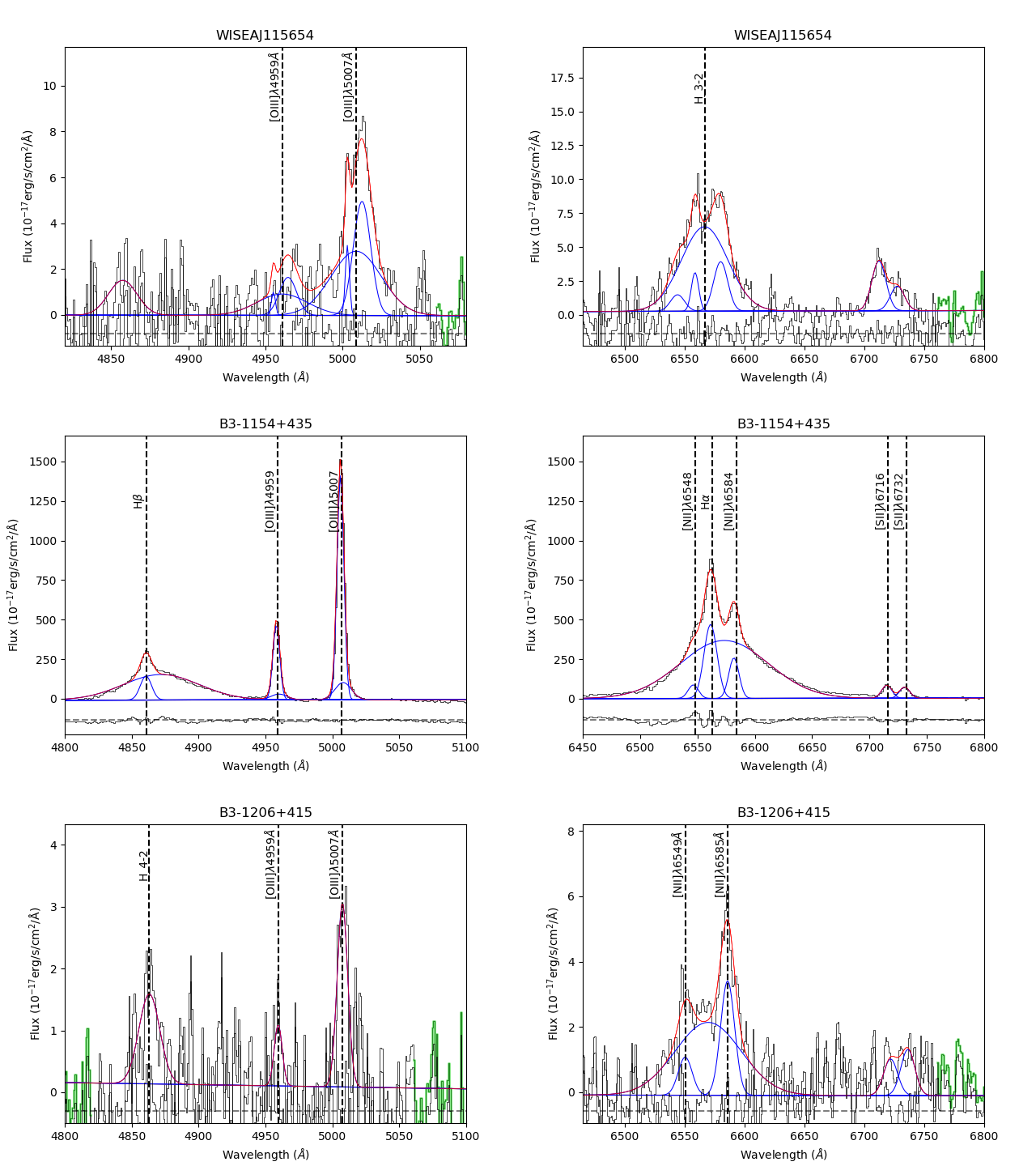} 

\caption{Integrated spectra fitting of the CSS and GPS sources with the residuals.   }
\label{fig:fitting-css}
\end{center}
\end{figure*}

\begin{figure*}
\begin{center}

\includegraphics[width=\textwidth]{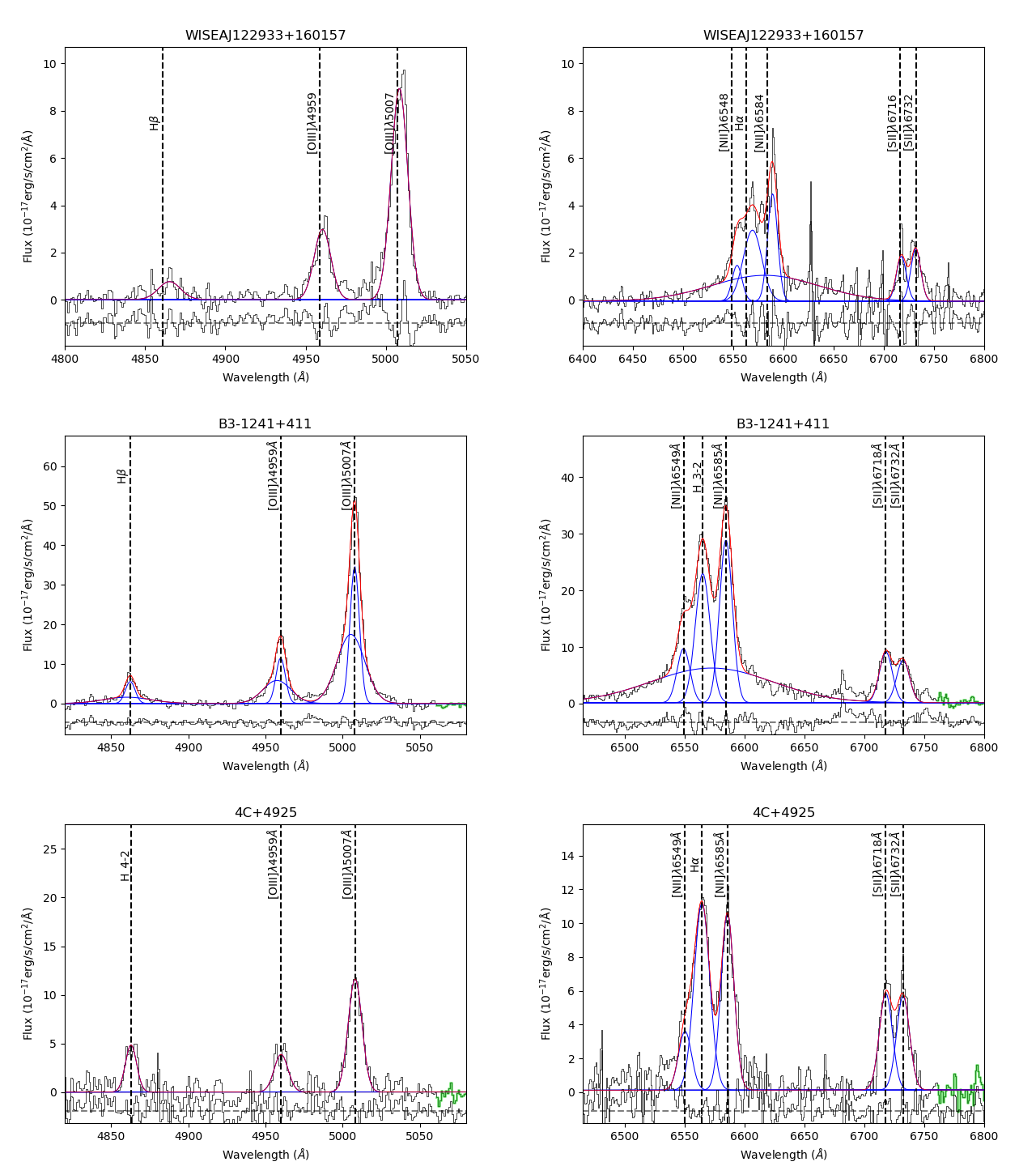}

\caption{Integrated spectra fitting of the CSS and GPS sources with the residuals.   }
\label{fig:fitting-css}
\end{center}
\end{figure*}

\begin{figure*}
\begin{center}

\includegraphics[width=\textwidth]{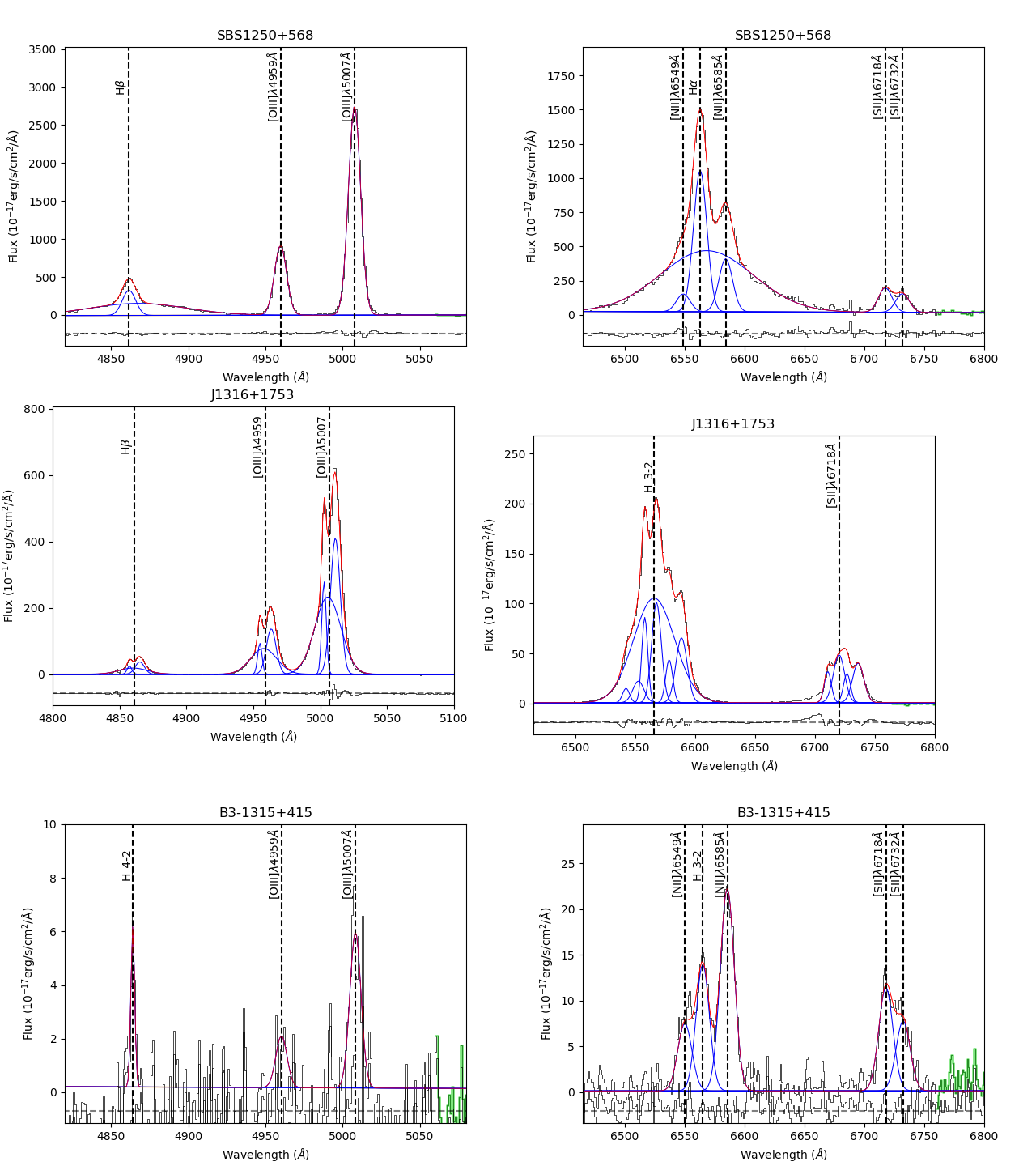}

\caption{Integrated spectra fitting of the CSS and GPS sources with the residuals.}
\label{fig:fitting-css}
\end{center}
\end{figure*}

\begin{figure*}
\begin{center}

\includegraphics[width=\textwidth]{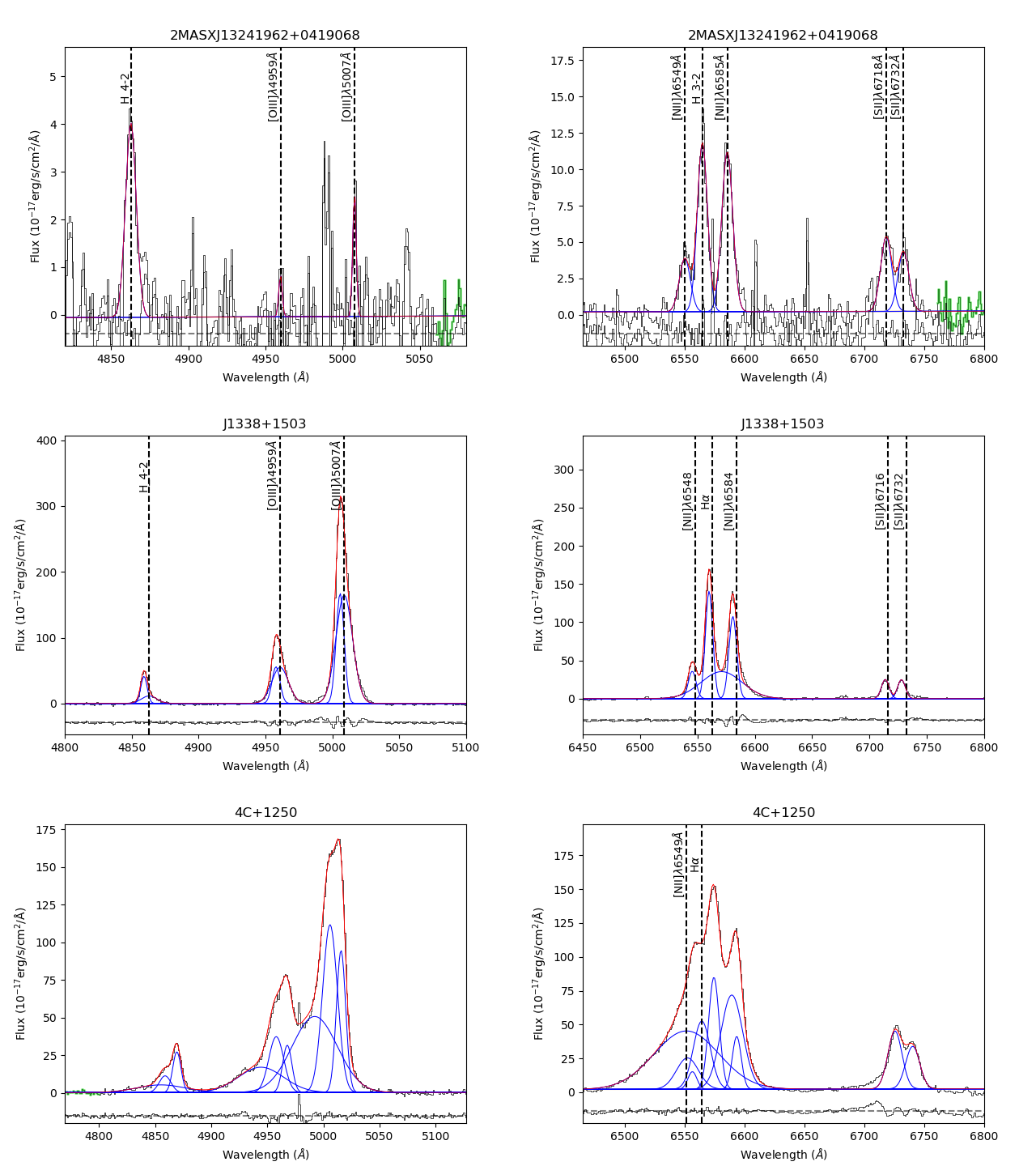}

\caption{Integrated spectra fitting of the CSS and GPS sources with the residuals.}
\label{fig:fitting-css}
\end{center}
\end{figure*}

\begin{figure*}
\begin{center}

\includegraphics[width=\textwidth]{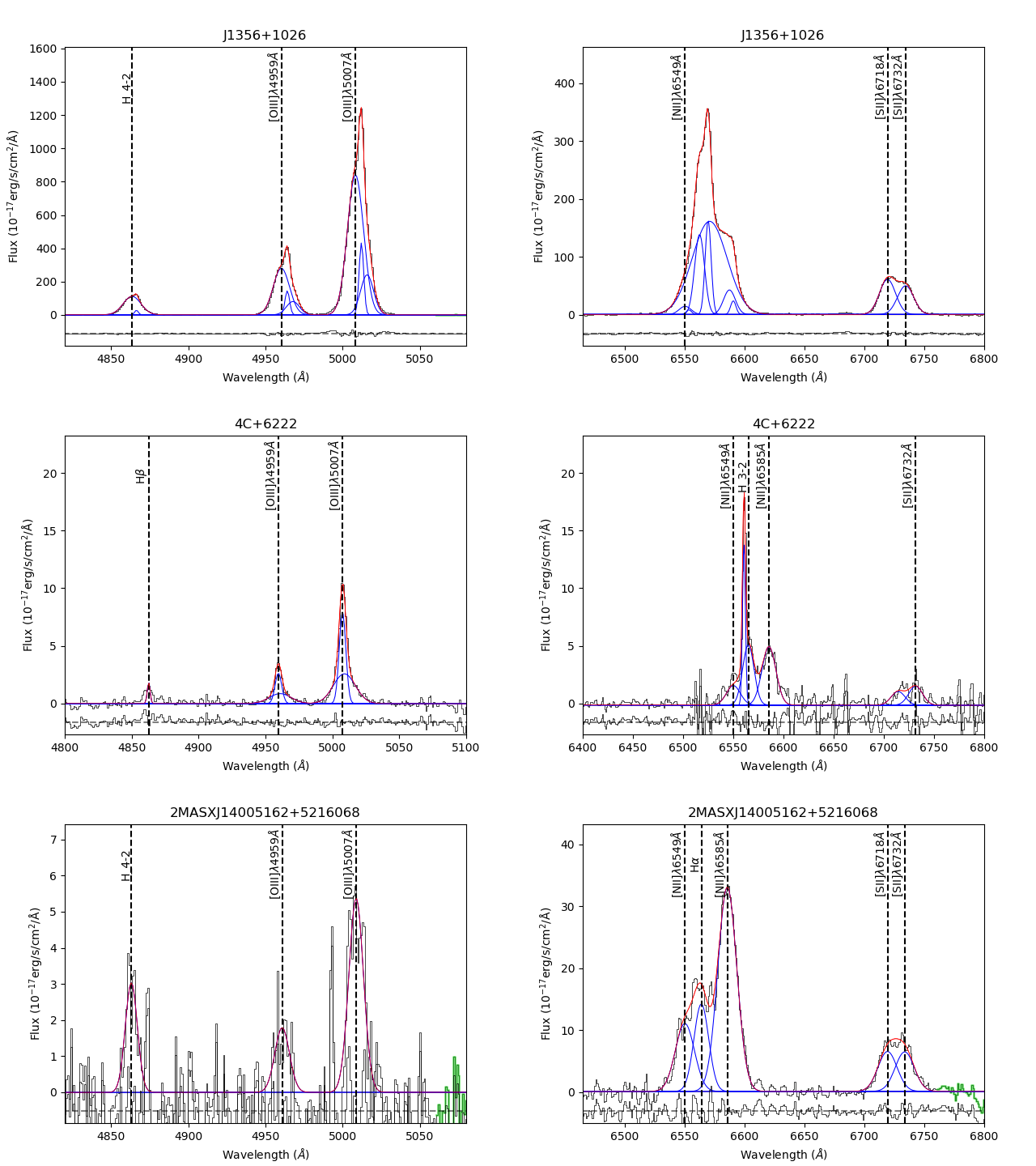}

\caption{Integrated spectra fitting of the CSS and GPS sources with the residuals.   }
\label{fig:fitting-css}
\end{center}
\end{figure*}

\begin{figure*}
\begin{center}

\includegraphics[width=\textwidth]{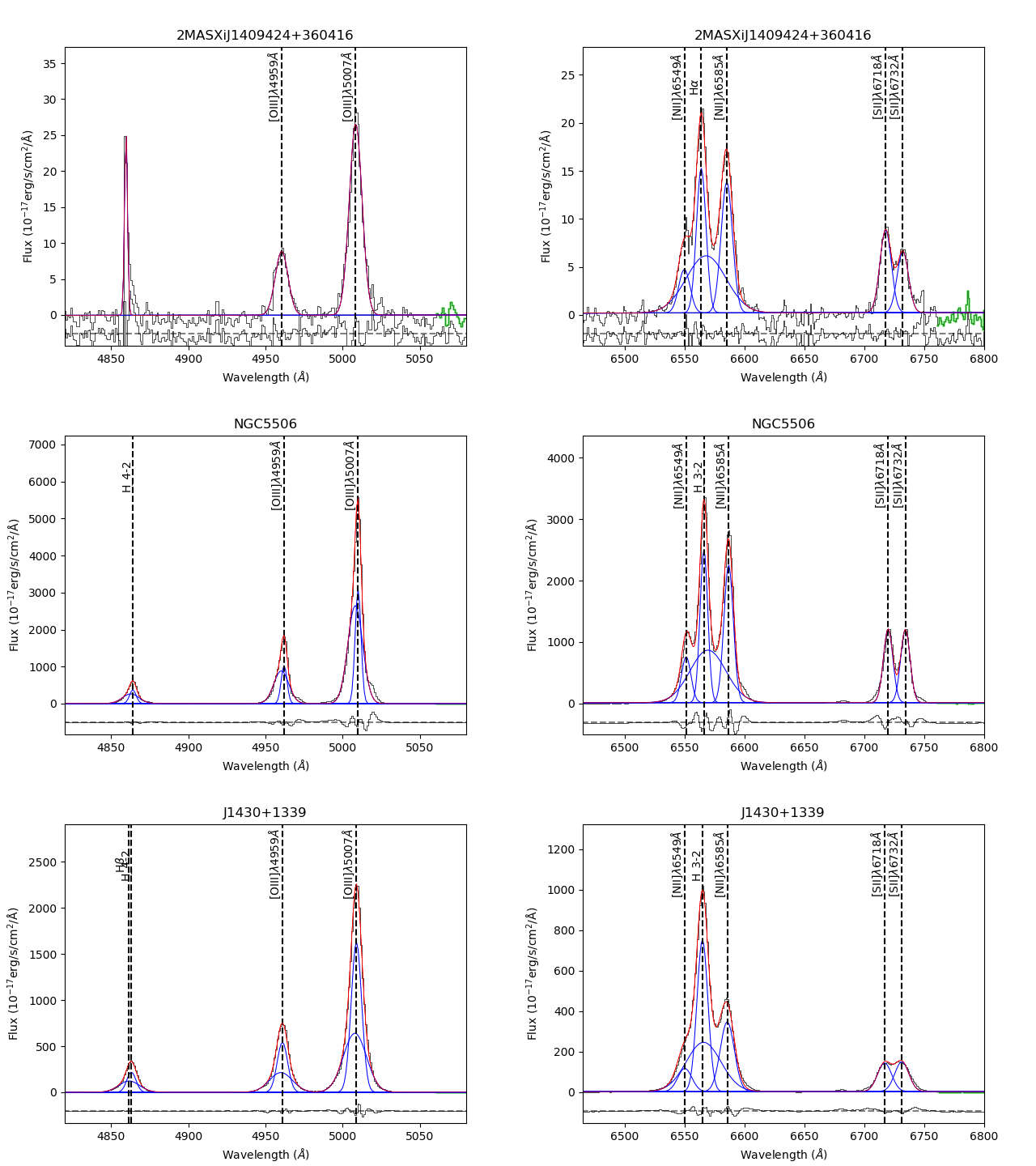}

\caption{Integrated spectra fitting of the CSS and GPS sources with the residuals. }
\label{fig:fitting-css}
\end{center}
\end{figure*}

\begin{figure*}
\begin{center}

\includegraphics[width=\textwidth]{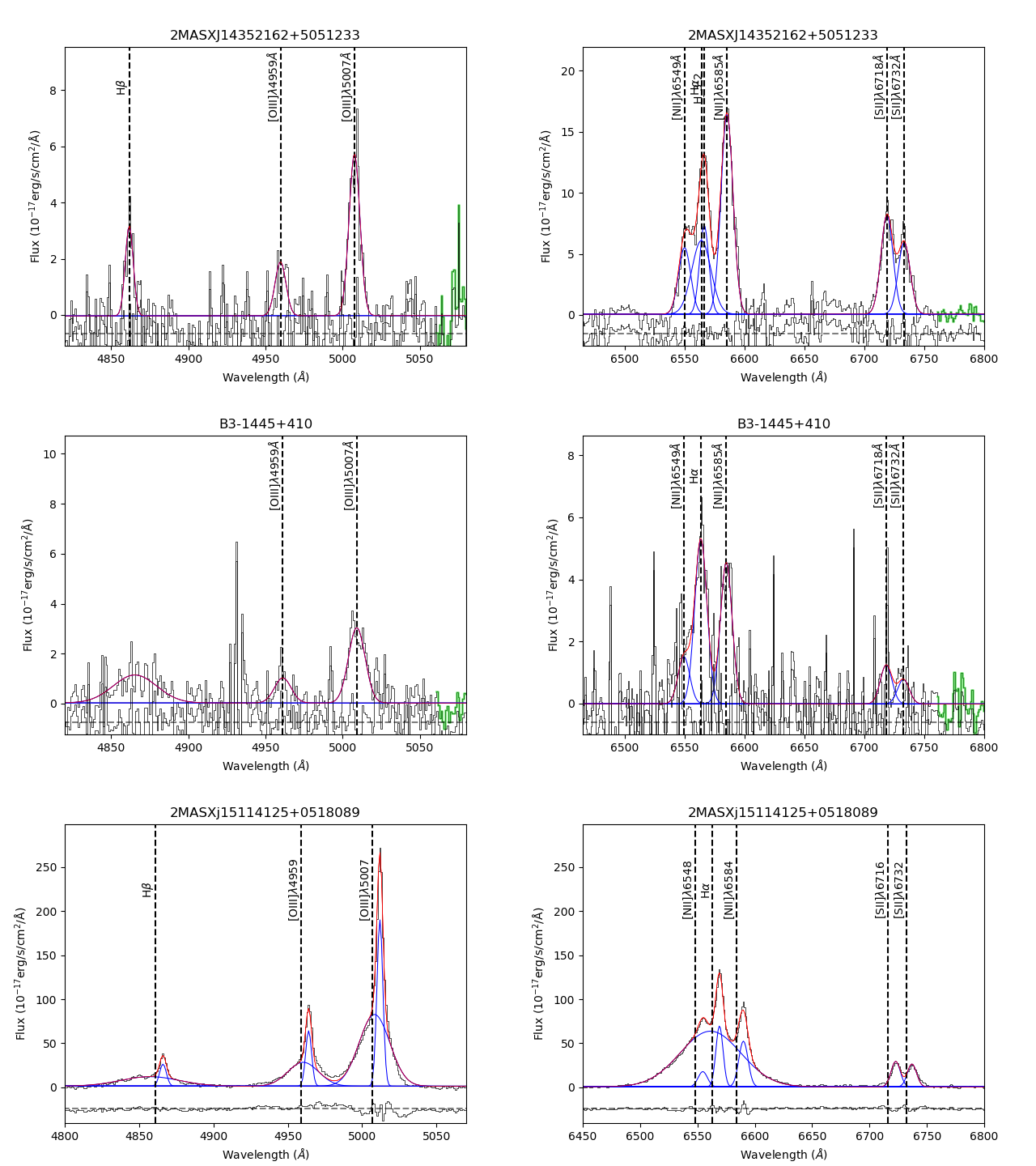}

\caption{Integrated spectra fitting of the CSS and GPS sources with the residuals. }
\label{fig:fitting-css}
\end{center}
\end{figure*}

\begin{figure*}
\begin{center}

\includegraphics[width=\textwidth]{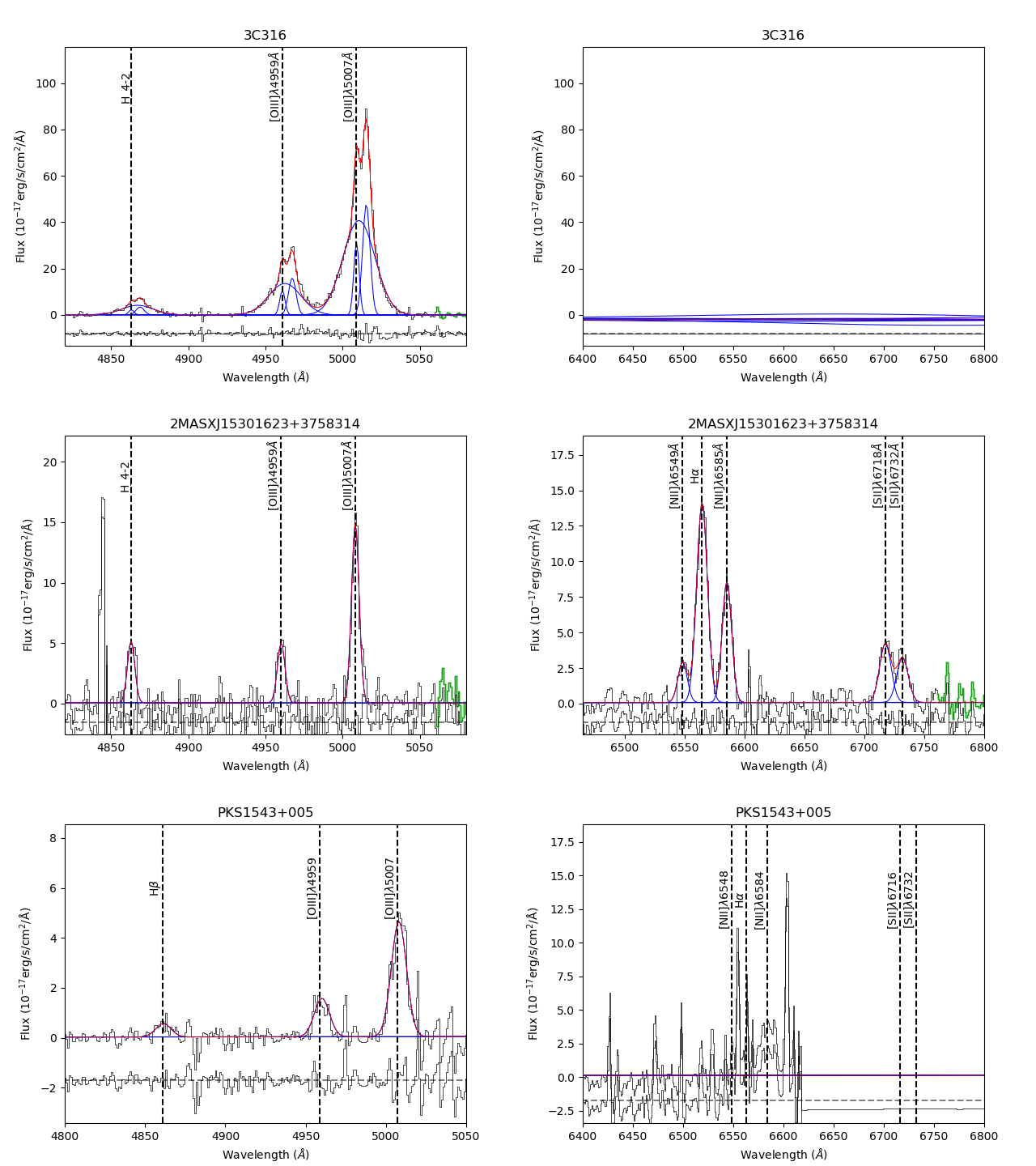}

\caption{Integrated spectra fitting of the CSS and GPS sources with the residuals. }
\label{fig:fitting-css}
\end{center}
\end{figure*}

\begin{figure*}
\begin{center}

\includegraphics[width=\textwidth]{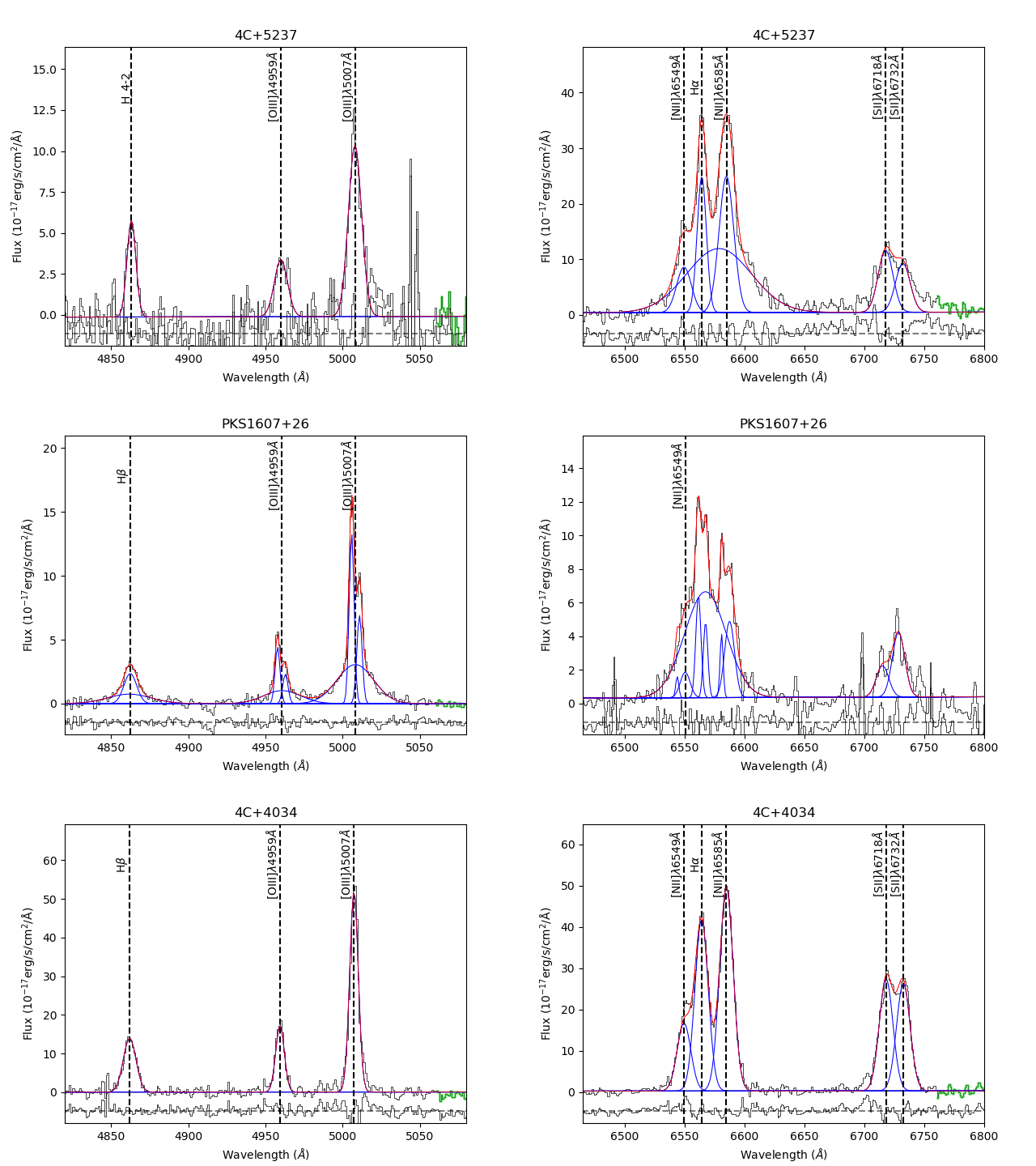} 

\caption{Integrated spectra fitting of the CSS and GPS sources with the residuals. }
\label{fig:fitting-css}
\end{center}
\end{figure*}

\begin{figure*}
\begin{center}

\includegraphics[trim={0 0 0 12cm},clip,width=\textwidth]{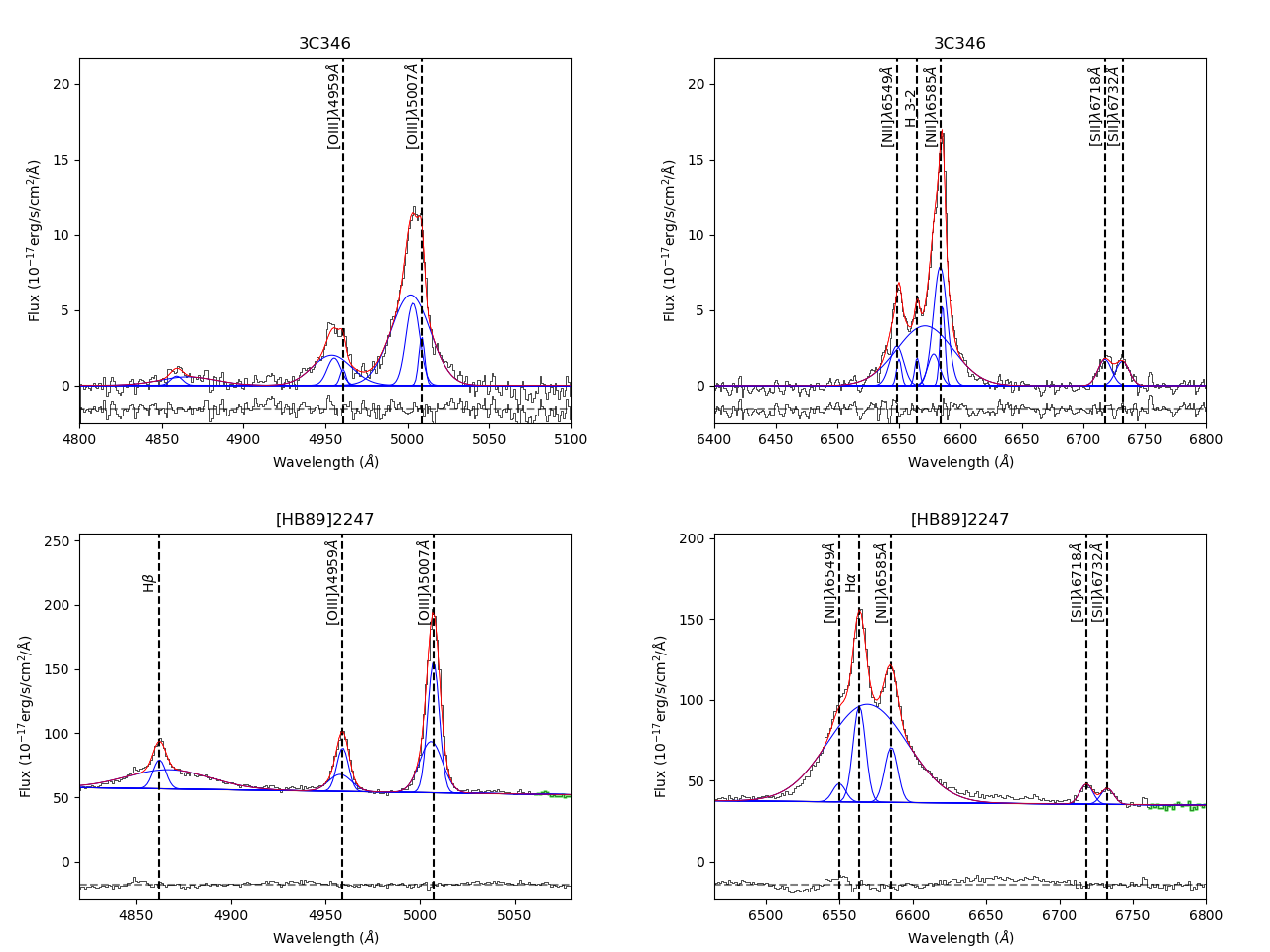} 

\caption{Integrated spectra fitting of the CSS and GPS sources with the residuals.   }
\label{fig:fitting-css2}
\end{center}
\end{figure*}

\end{document}